\documentclass[10pt,journal,compsoc]{IEEEtran}
\usepackage{graphicx} % Required for inserting images
\usepackage[export]{adjustbox}
\usepackage{multirow}
\usepackage{subcaption}
\usepackage{color}
\usepackage{hyperref}
\usepackage{etoolbox}

\def\fixme#1{\typeout{FIXED in page \thepage : {#1}}
\bgroup \color{red}{[FIXME: {#1}]} \egroup}
\def\reply#1{\typeout{REPLY in page \thepage : {#1}}
\bgroup \color{blue}{[REPLY: {#1}]} \egroup}
\def\new#1{\typeout{NEW in page \thepage : {#1}}
\bgroup \color{green}{[NEW: {#1}]} \egroup}

\DeclareRobustCommand{\&}{%
  \ifdim\fontdimen1\font>0pt
    \textsl{\symbol{`\&}}%
  \else
    \symbol{`\&}%
  \fi
}

\AtBeginEnvironment{quote}{\small}

\title{Analysis and Mitigation of Shared Resource Contention on Heterogeneous Multicore: An Industrial Case Study}
\author{Michael Bechtel, Heechul Yun \\
}

\begin{document}

% \keywords{Industrial Challenge, Real Time, SLAM, Microarchitectural DoS Attacks

\maketitle

\footnotetext[1]{\textit{Dr. Bechtel is currently with Garmin. This work was conducted while he was with the University of Kansas. E-mail: mgbechte@gmail.com}}
\footnotetext[2]{\textit{Dr. Yun is with the University of Kansas and is the corresponding author of this manuscript. E-mail: heechul.yun@ku.edu}}

\begin{abstract}

% Heterogeneous multicore systems are becoming increasingly used due to the improved performance they offer. This can be seen in the embedded and automotive domains, where heterogeneous platforms are incorporated into various autonomous systems such as self-driving cars. However, shared resource contention remains a significant challenge for these multicore platforms, with both academia and industry players working to tackle its impacts. 
%In fact, due to the significance of this problem, more wide-spread efforts have been introduced in recent years to better stimulate new innovations in combating shared resource contention.

In this paper, we present a solution to the industrial challenge put forth by ARM in 2022. We systematically analyze the effect of shared resource contention to an augmented reality head-up display (AR-HUD) case-study application of the industrial challenge on a heterogeneous multicore platform, NVIDIA Jetson Nano. We configure the AR-HUD application such that it can process incoming image frames in real-time at 20Hz on the platform. We use Microarchitectural Denial-of-Service (DoS) attacks as aggressor workloads of the challenge and show that they can dramatically impact the latency and  accuracy of the AR-HUD application. This results in significant deviations of the estimated trajectories from known ground truths, despite our best effort to mitigate their influence by using cache partitioning and real-time scheduling of the AR-HUD application. To address the challenge, we propose RT-Gang++, a partitioned real-time gang scheduling framework with last-level cache (LLC) and integrated GPU bandwidth throttling capabilities. By applying RT-Gang++, we are able to achieve desired level of performance of the AR-HUD application even in the presence of fully loaded aggressor tasks.

%We show that by combining dynamic LLC bandwidth throttling of the aggressor tasks and GPU bandwidth throttling of the head-pose estimation task, we can effectively ensure real-time performance of the AR-HUD application without resorting to over-provisioning the system. 

% We find that a certain type of DoS attack~\cite{bechtel2023cache}, when added to the system as aggressor workloads, is especially effective in increasing inaccuracies in the SLAM generated trajectory, and in increasing the execution times to a degree that it completely fails to meet the real-time requirements. 
% In doing so, we find that both CPU and GPU co-runners can degrade the performance of the critical Simultaneous Localization and Mapping (SLAM) algorithm such that its trajectory deviates from a known ground truth by several meters on average. 
% Furthermore, in the worst case, we show that a combination of CPU and GPU co-runners can prevent the SLAM algorithm from processing in real-time, to the point where only $\sim$26\% of input frames are processed. To mitigate this contention, we extend a gang scheduling framework to support LLC bandwidth throttling of best-effort tasks only. Using the extended framework, we can protect the SLAM algorithm such that it can process in real-time again, but find additional efforts for isolation from the GPU to be necessary.
    
\end{abstract}


%-------------------------------------------------------------------------
\section{Introduction} \label{sec:intro}

Heterogeneous multicore computing platforms are increasingly utilized in safety-critical cyber physical systems (CPS) as they can offer significant performance improvements while simultaneously meeting size, weight, and power (SWaP) constraints. 
% These platforms incorporate heterogeneous computing elements, such as GPUs, that can aid in the processing of complex machine learning algorithms. 
However, contention on shared microarchitectural resources, such as shared cache and main memory, between the computing elements in such a platform remains a significant challenge because it can impact the execution timings of critical real-time tasks and thus jeopardize the  safety of the CPS. Moreover, shared resource contention can also be intentionally induced by malicious actors with the goal of compromising the performance and safety of CPS. Such adversaries are known as microarchitectural Denial-of-Service (DoS) attacks~\cite{bechtel2019dos}, and are especially problematic for high-performance CPS that need to run multiple concurrent applications simultaneously on a single multicore platform. Given the recent trends towards  connected CPS, as can be seen in ARM's SOAFEE initiative~\cite{soafee} for the automotive industry, it is conceivable that such DoS attacks could be remotely deployed on future CPS.

Understanding and addressing shared resource contention in multicore has been of intense interest for both academia and industry in recent years.
%, so much so that more organized efforts to combat it have been organized in recent years. 
In particular, ARM issued an Industrial Challenge in 2022 to address the problem of shared resource contention~\cite{andreozzi2022industrial}. The challenge is centered around an augmented reality head-up display (AR-HUD) case-study for automotive applications. The case-study application is composed of two main components: a Visual Simultaneous Localization and Mapping (SLAM) task~\cite{ferrera2021ov} and a DNN-based driver head pose estimation task~\cite{hopenetlite}. The SLAM task is composed of three main threads, all of which run on the CPU, whereas the DNN-based head-pose estimation task (we henceforth refer to it as the DNN task) may utilize the GPU. The application represents a computationally intensive mixed-criticality real-time system that must leverage high-performance heterogeneous multicore embedded platforms. As such, the challenge seeks to find ways to analyze and optimize performance bounds of such critical real-time tasks even in the presence of ``aggressor tasks'', which may contend with the critical real-time task in accessing shared resources.

% 
%Of the two components, the SLAM task is of higher real-time priority, meaning its performance is of higher importance for the overall AR-HUD application. 

%\fixme{move this to related work} 
%In addition, SLAM algorithms are susceptible to performance loss due to timing variations~\cite{li2021chronos,li2022timing}. Due to this, we primarily focus on the Visual SLAM component in this paper.

In this paper, we first study the impact of shared resource contention to the performance of the AR-HUD case-study application of ARM's Industrial Challenge. Through our study, we aim to answer the following questions:
(1) Can we safely consolidate the two real-time tasks (the SLAM algorithm and head pose detection) in the AR-HUD case-study on a representative heterogeneous system-on-chip (SoC) processor and achieve required real-time performance (i.e., meeting the deadlines)? 
(2) Does shared resource contention between the two AR-HUD tasks impact the accuracy of the obtained position/trajectory estimates of the SLAM task?
(3) Can we guarantee a desired level of performance, in terms of both accuracy and latency, of the AR-HUD application in the presence of aggressor tasks---which may be maliciously designed to cause high shared resource contention---without excessive over-provisioning?

To answer these questions, we systematically conduct experiments on an NVIDIA Jetson Nano, a representative heterogeneous embedded multicore platform that features a quad-core ARM Cortex-A57 CPU and an integrated GPU. Our findings are as follows: 
We are able to configure the AR-HUD application such that it can process incoming image frames in real-time at 20Hz. However, we find that contention between the two real-time tasks does significantly impact the accuracy of the SLAM task, which results in significant deviations of the estimated trajectories from the ground truth even when it could process all input image frames in real-time at 20Hz. 
In addition, we find that cache bank-aware DoS attack~\cite{bechtel2023cache} is especially effective in impacting the accuracy and real-time performance of the AR-HUD application. 
Concretely, when the cache bank-aware DoS attack tasks are co-scheduled as best-effort (non-RT) tasks together with the real-time tasks of the AR-HUD application to fully load the system, the SLAM task fails to even generate the trajectory as it has to drop most of the incoming image frames due to increased latency caused by contention.
%observe up to $\sim$55X increase in absolute trajectory error (ATE) compared to the baseline. 
% Likewise, the head pose estimator can increase the deviation by $\sim$27X when run concurrently. \fixme{What do you exactly mean?}
%Most notably, though, when run alongside the both DoS attackers and the head pose estimator, we find that SLAM performance suffers significantly. 
%the SLAM task was unable to keep up with the input sensor data, and can only successfully process $\sim$26\% of the input frames, resulting in a complete system failure. 

%As such performance would most likely lead to system failure in a real-world scenario, we also explore mechanisms that can be utilized to address this interference.
%\fixme{need to assume readers are not aware of rt-gang, even what gang scheduling is}
%\reply{I added a brief descriptions of gang scheduling and RT-Gang, and a background section that goes into more detail.}
% We then explore a mitigation method to help protect the performance of the real-time tasks from the shared resource contention.
To address the challenge, we propose RT-Gang++, a partitioned real-time gang scheduling framework with iGPU and last level cache (LLC) bandwidth throttling capabilities. RT-Gang++ is based on ~\cite{ali2019rt} but extends its capabilities as follows: (1) add support for partitioned gang-scheduling to allow for multiple real-time gangs of different priorities to execute concurrently; (2) add support for LLC and iGPU bandwidth throttling to protect against contention on those shared resources. These additional capabilities are crucial to address the ARM industrial challenge problem. 

By employing RT-Gang++, we are able to safely consolidate the AR-HUD application on the Jetson Nano platform, even in the presence of malicious DoS attacks, and achieve desired real-time performance and accuracy. In addition, we also ported RT-Gang++ on a Raspberry Pi 4 platform, and evaluate its effectiveness.
For reproducible dissemination, we release the AR-HUD application setup, including our ROS2 port of OV$^2$SLAM, evaluation scripts, and RT-Gang++, as open-source~\footnote{\url{https://github.com/CSL-KU/ArmArHudChallenge}}.

The rest of this paper is organized as follows. Section~\ref{sec:bac-vslam}  describes the ARM Industrial Challenge problem. Section~\ref{sec:bac-dos} describes microarchitectural DoS attacks for the challenge. Section~\ref{sec:case} discusses our experimental setup. Section~\ref{sec:evaluation} presents our empirical evaluation of the challenge's case-study application. 
Section~\ref{sec:mitigation} presents a mitigation approach. Section~\ref{sec:results} presents the results. 
We review related work in Section~\ref{sec:related} and conclude in Section~\ref{sec:conclusion}.
% \section{ARM Industrial Challenge 2022} \label{sec:background}

% % In this section, we provide necessary background information on visual SLAM algorithms, and microarchitectural DoS attacks. % and real-time gang scheduling.
% %\fixme{background on RT-Gang is missing} \reply{I added an initial section for RT-Gang and real-time gang scheduling.}
% \fixme{explain the challenge and its case study application. e.g., ARM created the challenge because of the increased importance of consolidating high-performance mixed criticality applications in cyber-physical systems, particularly those in the automotive domain. we briefly introduce the suggested case-study application. }
% \reply{I added a brief section on the challenge motivation.}

\section{ARM Industrial Challenge 2022: Augmented Reality Head-Up Display (AR-HUD) Application} \label{sec:bac-vslam}

Addressing the impacts of shared resource contention is of critical importance for many high-performance CPS, such as those in the robotics and automotive fields. This is especially the case given the increased importance of consolidating high-performance mixed criticality applications in CPS. To stimulate further research on this topic, ARM introduced an Industrial Challenge in ECRTS 2022. The challenge presented an augmented reality head-up display (AR-HUD) application in the automotive context as a case-study. As an advanced driver assistance system (ADAS), this application provides additional alerts and notifications to the driver of a vehicle in real-time. In particular, these alerts are overlaid on real-world objects using augmented reality (AR) technology. For the suggested AR-HUD application, it is mainly comprised of two components: a Visual SLAM task, and a head pose estimation task. We now briefly introduce and discuss both AR-HUD components.

For many autonomous cyber physical systems (CPS), localization and 3D map generation are important steps for real-world performance. Increasingly, many CPS employ Simultaneous Localization and Mapping (SLAM) algorithms to perform both operations in a single step. In a SLAM algorithm, input sensor data is received and utilized to both estimate a system's current position in, and generate/update a 3D map of a given environment. Vision (camera) and range sensors such as LIDARs, lasers and sonars can be used for SLAM. The category of SLAM algorithms that utilize vision has come to be known as \textit{Visual SLAM}.

In the ARM industrial challenge, the OV$^2$SLAM algorithm~\cite{ferrera2021ov} is suggested as part of the AR-HUD case study~\cite{andreozzi2022industrial}. OV$^2$SLAM is a Visual SLAM algorithm that is geared toward real-time applications and emphasizes processing time in addition to SLAM performance. It is composed of four main components with each one being assigned to a separate thread:
\begin{enumerate}
    \item The \textit{Front-End} thread performs real-time pose estimation of the camera sensor. It is also responsible for creating the keyframes used to generate 3D maps of surrounding environments, but does not create a keyframe for every given frame. Note that this thread runs for every input frame that is received, meaning that it is a periodic task in nature. For our purposes, we target a per-frame deadline of 50 ms as the input datasets we use in our evaluations playback data at a frequency of 20 Hz.
    \item The \textit{Mapping} thread uses keyframes generated in the \textit{Front-End} to generate new 3D map points. It primarily does this by performing triangulation on the keyframes. Then, if a new keyframe has not arrived, it will also perform local map tracking in order to minimize drift. Unlike the \textit{Front-End}, the \textit{Mapping} thread is aperiodic as it is event-driven and only runs when a new keyframe is generated.
    \item The \textit{State Optimization} thread performs two main operations. First, it runs a local bundle adjustment (BA) to refine camera pose estimations. Second, it runs a keyframe filtering pass that prevents redundant keyframes from being processed in future BA operations. Note that this thread is also aperiodic as it relies on input from the \textit{Mapping} thread, meaning that it is also event-driven.
    \item The \textit{Loop Closer} thread performs an online bag-of-words (BoW) operation to detect loop closures in a system's given trajectory. However, we do not employ this thread in our case study as it is not necessary for the target AR-HUD application~\cite{andreozzi2022industrial}.
\end{enumerate}
Note that only the \textit{Front-End} thread runs for every input frame that is fed to OV$^2$SLAM. The remaining threads will then only run when necessary, such as when a new keyframe is created.

By default, the OV$^2$SLAM algorithm can be run in one of three different modes: \textit{accurate}, \textit{fast}, and \textit{average}. The \textit{accurate} mode of operation performs all four steps described above, including Loop Closure, and is intended to maximize accuracy while still maintaining a control frequency of 20 Hz. On the other hand, the \textit{fast} mode of operation instead sacrifices some accuracy so that it can operate at a much faster 200 Hz control frequency. To achieve this, the \textit{fast} version uses a faster (but less accurate) keypoint detection algorithm and does not perform the Loop Closure step. The \textit{average} version then operates in between the other two versions performance-wise. In other words, it runs at a control frequency between 20 and 200 Hz, and achieves accuracy worse than the \textit{accurate} version but better than the \textit{fast} version. Like the \textit{accurate} version, though, the \textit{average} version also performs Loop Closure. The Industrial Challenge suggests to use the \emph{fast} version for its superior real-time performance and good accuracy. 

For the AR-HUD application, it is also important that the ADAS alerts provided to the driver are displayed in a way that matches the driver's viewpoint. To achieve this, the head pose of the driver can be estimated so that the AR display can be corrected as necessary. As such, the AR-HUD application employs a head pose estimation task for its second component. %One example of a high performance head pose estimator is that of the HopeNet DNN~\cite{ruiz2018hopenet}. However, 
The Industrial Challenge suggests to use the HopeNet-Lite head pose estimator~\cite{hopenetlite}, as it can run in real-time on many embedded heterogeneous multicore platforms. We further discuss this component in our evaluation setup. % In our case, we employ the suggested HopeNet-Lite DNN model~\cite{hopenetlite}. A lightweight version of the original HopeNet model~\cite{ruiz2018hopenet}, HopeNet-Lite employs the ShuffleNet V2 network~\cite{ma2018shufflenet} and is implemented using the PyTorch framework~\cite{paszke2019pytorch}. In our testing, we found that an instance of the HopeNet-Lite model could process input frames at $\sim$35 ms per frame. As such, we use an initial control frequency of 25 Hz for the head pose estimation task in our case study. In terms of priority, because the head pose estimation task is listed as \textit{high-priority} in the AR-HUD application, we assign all HopeNet-Lite model instances to a real-time priority of 1 so that they are of lower priority than the SLAM task. In addition, we pin the HopeNet-Lite task to a CPU core distinct from the SLAM task cores, Core 3, so that none of its required CPU operations (e.g. CUDA kernel launch, etc.) run on the same core as an OV$^2$SLAM thread.

\section{Microarchitectural Denial-of-Service (DoS) Attacks} \label{sec:bac-dos}

% \fixme{the challenge envision the existence of ``aggressor workloads'', which may be co-scheduled with the AR-HUD application described above and content on the shared sources. For the aggressor workloads, we use micro-architectural DoS attacks described in literature. These DoS attacks target various micro-architectural resources in multicore and were shown to be highly effective in causing massive execution dealy on co-scheduled tasks even when they are running on dedicated cores and dedicated cache partitions. In this work, we wanted to know the there impacts on the AR-HUD case study application. The DoS attacks we evaluated are as follows.}
% \reply{I added the aggressor workload motivation and moved the listing of the DoS attacks from the case study section to here.}

As part of the Industrial Challenge, ARM also envisioned the presence of \textit{aggressor workloads} in the AR-HUD case study~\cite{andreozzi2022industrial}. These aggressor workloads may be co-scheduled alongside the AR-HUD application and may contend for shared resources. 

For our aggressor workloads, we employ microarchitectural denial-of-service (DoS) attacks that have been described in literature~\cite{valsan2016taming,bechtel2019dos,bechtel2023cache}. These DoS attacks target various microarchitectural resources in multicore platforms (e.g. LLC, DRAM) and can cause significant execution time delays to cross-core real-time tasks, even if they run on dedicated cores and have dedicated LLC partitions. In this work, we want to know the impacts such DoS attacks can have on the AR-HUD application. The specific DoS attacks we employ in our evaluations are as follows:
\begin{itemize}
    \item The \textit{bandwidth} benchmark from the IsolBench suite~\cite{valsan2016taming}. This benchmark is designed to perform continuous accesses to a target shared resource (e.g. LLC or DRAM) in a sequential manner. To be more specific, it performs sequential accesses over a 1D array at a cache line granularity (i.e. all accesses are 64B apart). We refer to DoS attacks based on this benchmark as \textit{Bw}.
    \item The \textit{latency-mlp} benchmark from the IsolBench suite~\cite{valsan2016taming}. Much like the \textit{bandwidth} benchmark, \textit{latency-mlp} continually accesses a target resource but differs in its access pattern due to its pointer chasing nature. Namely, it performs random accesses over multiple parallel linked lists (PLL). We refer to DoS attacks based on this benchmark as \textit{PLL}.
    \item The cache bank-aware attacks from~\cite{bechtel2023cache}. Much like memory-aware attacks~\cite{bechtel2021memory}, these attacks are based on the \textit{PLL} attacks above but are modified to only access a specific cache bank in order to generate maximum cache bank contention in accessing the LLC.
    % In addition, these attackers can be configured to perform sequential or random access patterns, much like the \textit{Bw} and \textit{PLL} attackers discussed above. However, we only employ the \textit{PLL} based attacks as they were shown to be more effective. 
    We refer to this attack as \textit{BkPLL}.
\end{itemize}
Furthermore, these DoS attacks can be configured in two additional facets. First, they can all be configured to perform either read or write accesses. As such, we test DoS attacks of both access variations in our testing. Second, as mentioned above, the attackers can be configured to access the LLC or DRAM, so we employ separate DoS attacks targeting each shared resource. Note that we configure the \textit{BkPLL} attacks to only access the LLC, as they are specifically designed for that resource. Putting it altogether notation wise, we use the following naming convention for DoS attacks:

\begin{quote}
    $<$\textit{DoS attack type}$><$\textit{access type}$>$($<$\textit{target resource}$>$) 
\end{quote}

Where \textit{DoS attack type} is one of the attacks from the above list, the \textit{access type} is either read or write, and the \textit{target resource} is either the LLC or DRAM. For example, an instance of the \textit{Bw} attack that performs read accesses targeted to the LLC would be referred to as \textit{BwRead(LLC)}. In total, we employ 10 different DoS attacker tasks in our evaluation.
\section{Experiment Setup} \label{sec:case}

In this section, we describe the experimental setup for our case study of the AR-HUD application. 

% \subsection{Hardware Platform}
{\bf Hardware Platform:} For the hardware platform, we use an Nvidia Jetson Nano platform, which equips a quad-core Cortex-A57 cores with each core having its own private L1 instruction and data caches and all cores sharing a global L2 cache. Table~\ref{tbl:platform} shows the basic characteristics of the platform. 
%Note that we primarily focus on the results from the Jetson Nano as it is more representative for the target AR HUD application. Likewise, we employ the Pi 4 to supplement the behavior and results we observe on the Jetson Nano.

\begin{table}[htp]
  \centering
  % \begin{adjustbox}{width=.49\textwidth}
  \begin{tabular}{|c|c|c|}
    \hline
    Platform                & Nvidia Jetson Nano \\ 
    \hline
    SoC                     & Tegra X1  \\ 
    \hline
    \multirow{1}{*}{CPU}    & 4x Cortex-A57 @ 1.43GHz   \\      
                            % & out-of-order       \\ 
                            % & 1.43GHz            \\ 
    \hline
    GPU                     & 128-core Maxwell \\
    \hline
    % Private L1 Cache        & 48KB(I)/32KB(D)    \\
    Shared LLC (L2)         & 2MB (16-way) \\ 
   %  \hline 
   % L2 (LLC) Bank Bits       & 4, 5, 6            \\
    \hline
    Memory (Peak B/W)       & 4GB LPDDR4 (25.6 GB/s) \\
    \hline
  \end{tabular}
  % \end{adjustbox}
  \caption{Nvidia Jetson Nano hardware specifications.}
  \label{tbl:platform}
\end{table}

%\subsection{Application Setup}
{\bf Application Setup:}
% which is the latest version that is officially supported by Nvidia. 
As discussed in Section~\ref{sec:bac-vslam}, the AR-HUD application is comprised of two main components: the OV$^2$SLAM Visual SLAM task and the HopeNet-Lite head pose estimation task. 
% In our evaluations, we primarily focus on the OV$^2$SLAM task as it is a high-criticality and higher priority task for the performance of the AR-HUD application as per~\cite{andreozzi2022industrial}. 
%As such, we begin by describing our experimental setup of the OV$^2$SLAM task itself.

For the SLAM task, we use the \textit{fast} setting of OV$^2$SLAM, as recommended in the Industrial Challenge~\cite{andreozzi2022industrial}, which is comprised of three threads: \textit{Front End}, \textit{Mapping}, and \textit{State Optimization}. The \textit{Front End} thread is invoked whenever the camera provides a new image frame, which is at a fixed rate of 20Hz. The \textit{Mapping} and \textit{State Optimization} threads are invoked conditionally when the \textit{Front End} thread generates a new key frame (See Section~\ref{sec:bac-vslam} for  details.)

As the SLAM task is \textit{real-time critical}~\cite{andreozzi2022industrial}, we use the Linux \texttt{SCHED\_FIFO} real-time scheduler and assign it a real-time priority of 2. Furthermore, we assign all three threads of the SLAM task onto two CPU cores, Core 0 and 1, which we experimentally determined to be sufficient when they run in isolation. Note that the maximum observed per-core CPU utilization of the SLAM threads is less than 70\%, meaning there is additional slack that can potentially be used by best-effort tasks.

As for input data for the SLAM task, we use the five Machine Hall (MH) scenarios of the EuRoC dataset~\cite{burri2016euroc}. Note that the EuRoC dataset includes visual-inertial data of a micro aerial vehicle (MAV), which includes stereo images, IMU measurements, and accurate motion and structure ground-truth data. In the MH scenarios, the data were collected while traversing through an indoor environment populated with various machinery with a varying degree of complexities, with the MH01 scenario being the easiest one for SLAM and the MH05 scenario being the hardest one.
%Note that the MH scenarios are ordered in terms of increasing difficulty, with the MH01 scenario being the easiest one for a SLAM application to map and MH05 being the hardest one. 
% We mainly include the results obtained in the MH01 scenario due to space consideration but all five scenarios show similar trends. 

The image frames from the dataset are fed to the SLAM task through an instance of \texttt{rosbag2}~\cite{rosbag2}, a tool that plays back datasets recorded in ROS bag files, which was running on Core 2 with a real-time priority of 2 such that it is not delayed by any best-effort tasks. 
Note that all MH datasets playback input data at a frequency of 20Hz.
The observed CPU utilization of \texttt{rosbag2} is about $\sim$5-10\% of the Core 2.

\begin{figure}[htp]
    \centering
    \includegraphics[width=0.49\textwidth]{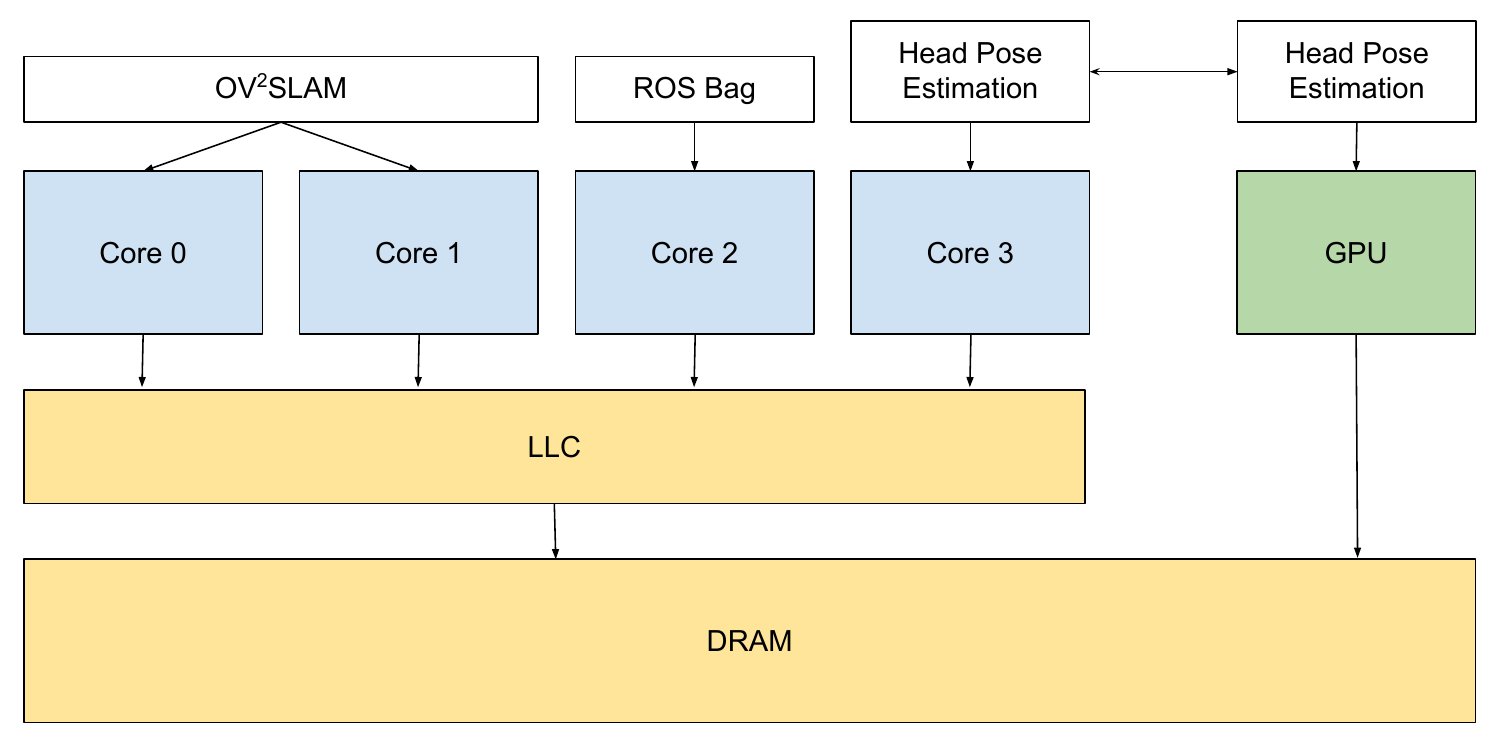}
    \caption{ Tasks to core assignments of the AR-HUD case-study on Jetson Nano. }
    \label{fig:case-study-setup}
\end{figure}

For the head pose estimation task, we use the recommended HopeNet-Lite DNN model~\cite{hopenetlite}, which is a lightweight version of the original HopeNet model~\cite{ruiz2018hopenet}. %  and uses the ShuffleNet V2 model~\cite{ma2018shufflenet} as its backbone. 
Note that the Jetson Nano mainly processes the HopeNet-Lite model on the GPU but a single CPU core (Core 3) is also used to launch the GPU kernel and monitor its progress. 
% The Raspberry Pi 4 instead solely runs the HopeNet-Lite model on its own CPU core 3.
% The HopeNet-Lite model could process input frames at $\sim$35 ms per frame when running in isolation on the Jetson Nano platform, and at $\sim$300 ms on the Raspberry Pi 4.
On the Jetson Nano we configure the DNN task to run periodically at the same 20Hz rate as the SLAM task and assign it a real-time priority of 1 as the task is determined to be a \textit{non-critical, high priority task}~\cite{andreozzi2022industrial}. 
% As such, we use an initial control frequency of 20 Hz for the head pose estimation task in our case study. 
% In terms of priority, because the head pose estimation task is listed as \textit{high-priority} in the AR-HUD application, we assign all HopeNet-Lite model instances to a real-time priority of 1 so that they are of lower priority than the SLAM task. 
In addition, we pin the HopeNet-Lite task to a CPU core distinct from the SLAM task cores, Core 3, so that none of its required CPU operations (e.g. CUDA kernel launch, etc.) are interfered with by the OV$^2$SLAM threads.

Figure~\ref{fig:case-study-setup} gives a visual representation of the setup we use and how we assign the AR-HUD tasks to CPU cores on the target platform. 
In addition,
Table~\ref{tbl:case-study-threads} shows the task/thread/core mapping and real-time scheduling parameters of all real-time tasks in the AR-HUD case study. 
%We next discuss the specific DoS attacker tasks that we employ and evaluate in our testing. 

\begin{table}[htp]
  \centering
  % \begin{adjustbox}{width=0.49\textwidth}
  \begin{tabular}{|c|c|c|c|c|}
    \hline
    \multirow{2}{*}{Task} & \multirow{2}{*}{Thread} & \multirow{2}{*}{Core(s)} & RT & Rate  \\ %  \multirow{2}{*}{Responsibilities}
    & & & Priority & (Hz) \\
    \hline
    \multirow{3}{*}{OV$^2$SLAM}  & Front-End & \multirow{3}{*}{0,1} & \multirow{3}{*}{2} & 20 \\ % RT pose estimation and new keyframe generation 
    \cline{2-2} \cline{5-5}
    & Mapping & & & - \\ % Generate 3D map points 
    \cline{2-2} \cline{5-5}
    & State Optimization & & & - \\ % Refine estimation and filter keyframes 
    \hline
    ROS Bag & -  & 2 & 2 & 20 \\ % Playback data used as input to OV$^2$SLAM 
    \hline
    Head Pose Est. & - & 3,GPU & 1 & 20 \\ % Estimation head pose of the driver in a car
    \hline
  \end{tabular}
  % \end{adjustbox}
  \caption{Real-time tasks/threads/core mapping and scheduling parameters in the AR-HUD case study on the Jetson Nano. Note that all real-time tasks are scheduled using the \texttt{SCHED\_FIFO} real-time scheduler 
  %, and (2) the Head Pose Estimation task only runs on CPU Core 3 on the Raspberry Pi 4. 
  and a bigger priority value indicates a higher real-time priority. }
  %\fixme{We need to tell that in this setup, the overall system utilization is about 60(?)\% or so and so there's plenty of slack that can potentially be utilized to run best-effort tasks for which we run DoS attacks} \reply{I added a few lines discussing this in the software setup section. Also, I double checked and the SLAM does use up to 70\% core utilization, but I never saw it go above that.}
  \label{tbl:case-study-threads}
\end{table}

% Table~\ref{sec:bac-vslam} shows the task/thread/core mapping and the real-time scheduling parameters of the real-time tasks of the AR-HUD case study.  

% \fixme{Add figure of experimental setup?} \fixme{yes, it show clearly show how many CPU threads are needed for slam and pose estimation. we also need to clarify our task-core assignment. } \reply{I added a basic figure of the core assignment and platform setup.}

% \subsection{Operating System Setup}
{\bf Operating System Setup:}
For the Jetson Nano's operating system we run Ubuntu 18.04 with Linux kernel 4.9, which is patched with PALLOC~\cite{yun2014palloc} to support LLC partitioning. PALLOC exploits virtual memory page translations to enforce page allocations to specific page colors.
With PALLOC, we partition the LLC into four equally sized partitions (colors) and perform a 2 by 2 split of those partitions. Namely, the OV$^2$SLAM algorithm gets two LLC partitions, and all other tasks share the remaining two cache partitions. 
Note that all best-effort tasks---those that are scheduled using Linux's default CFS scheduler---also share the latter two cache partitions in order to minimize any performance impact to the SLAM task, which is real-time critical. 
Note that, in PALLOC, tasks---not cores---can be mapped to any cache partitions.

\section{Analyzing the Effects of Shared Resource Contention} \label{sec:evaluation}

In this section, we evaluate the impact of shared resource contention on the performance of the OV$^{2}$SLAM algorithm.

\subsection{Impact of Co-scheduling DoS Attacks}~\label{sec:eval-dos}

\begin{figure*}[htp]
  \centering
  \includegraphics[width=0.73\textwidth]{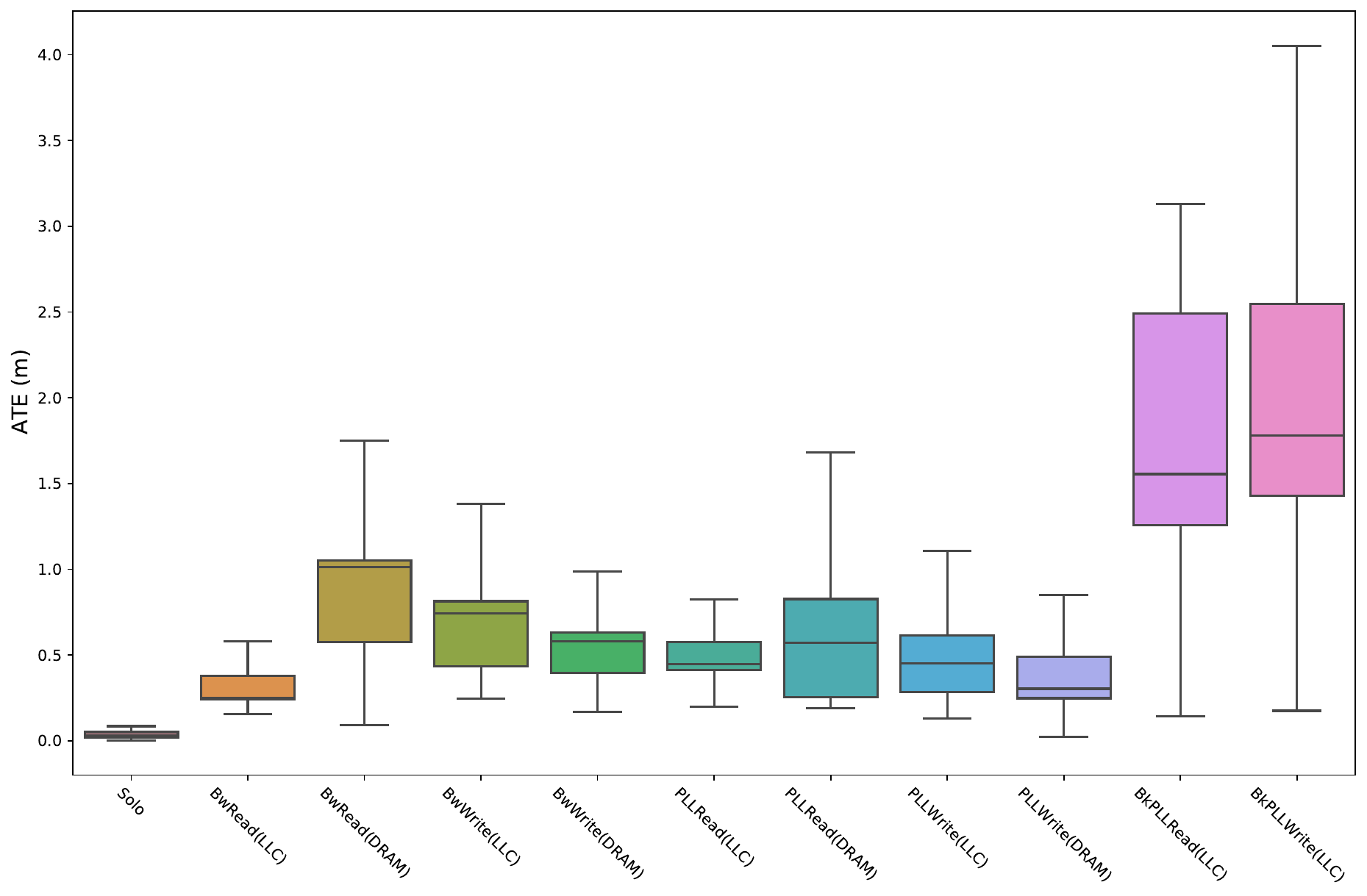}
  \caption{ Impact of DoS attacks on the Absolute Trajectory Error (ATE) of the OV$^2$SLAM generated trajectory.} %  \fixme{font is too small}
  \label{fig:dos-boxplot}
\end{figure*}

In this experiment, we evaluate the impacts of DoS attack co-runners and whether they are effective in degrading OV$^2$SLAM performance in a given scenario. 
Note that we do not execute the HopeNet-Lite DNN task in this experiment in order to focus on SLAM performance and its sensitivity to DoS attacks.

The experiment setup is as follows: We first run an instance of the OV$^2$SLAM task using the MH01 scenario in the EuRoC dataset~\cite{burri2016euroc}. Once finished, we calculate the algorithm's \textit{Absolute Trajectory Error (ATE)} relative to the known ground-truth trajectory~\cite{burri2016euroc}. 
We then repeat the experiment but with instances of a DoS attacker on all four available cores. 
We again calculate the ATE and compare it to the solo case to determine whether the attackers had any noticeable impact.

Note that we run the DoS attacks as best-effort tasks (scheduled using the CFS scheduler) while run the OV$^2$SLAM task as a real-time task (using the \texttt{SCHED\_FIFO}). Because Linux strictly prioritizes real-time tasks over best-effort ones, the DoS attack tasks can only be executed on cores which are not executing any of the RT tasks. 
% For the attacking tasks, we evaluate all DoS attacks described in Section~\ref{sec:bac-dos}. In each DoS attack case, we run an attacker task instance on each of the available CPU cores. For the Jetson Nano, this means we have four concurrent DoS attacker tasks. Unlike all of the AR-HUD tasks, we run all DoS attackers as best-effort tasks using Linux's default CFS scheduler with a default nice value, in order to prevent any performance loss due to scheduler conflicts. 
In other words, whenever the threads of the SLAM task become ready, they immediately preempt any DoS attacker tasks. Note also that, as mentioned earlier, the DoS attack tasks are assigned to a separate LLC cache partition from the SLAM task. This separation minimizes any negative effect of co-scheduling as the DoS attack tasks cannot evict cache-lines of the SLAM task. Therefore, any observed delays are not attributable to CPU scheduling or cache space contention. 
%Note that we do not mix instances of DoS attackers, we only employ one type of DoS attacker at a time. While investigation into the impacts of different DoS attacker combinations could be insightful, we leave such analysis to future work.

Figure~\ref{fig:dos-boxplot} shows a boxplot of the OV$^2$SLAM ATE collected over the entire duration of the MH01 dataset, both alone (Solo) and alongside each of the tested DoS attacks (the rest in the X-axis). 

Firstly, note that all of the tested DoS attacks cause significant negative impacts to the tracking performance, resulting in ATE increase of up to 4.0 meters. This occurs despite the fact that the DoS attacks cannot  preempt the SLAM task or evict its cache-lines. 
This is because there are many other shared hardware resources that can impact execution timing in modern multicore. These shared hardware resources include DRAM bandwidth~\cite{yun2013rtas,bechtel2021memory}, DRAM bank~\cite{yun2014palloc}, cache internal buffers/queues~\cite{valsan2016taming,bechtel2019dos}, and cache bank~\cite{bechtel2023cache}. The DoS attacks in the X-axis are designed to induce maximum contention in those shared resources. 
Note that, for this scenario, ATEs of $\sim$0.3 or more indicate significant deviations from the ground truth, which could potentially cause failure (e.g., a crash) in the real-world~\cite{li2022timing}. 

%Even in the best case, the \textit{BkBwWrite(LLC)} attacks still increased the median ATE by $\sim$6.8X, going from a solo ATE of $\sim$0.035 to $>$0.24. 
%Note that 
% This is because the use of real-time scheduling does not prevent shared resource contention caused by the DoS attackers running on different cores.
%Based on this, we believe that the performance losses caused by all of tested DoS attacks are noteworthy and could potentially cause system failures in a real-world CPS.
In particular, we observe that the cache bank-aware DoS attacks recently proposed in~\cite{bechtel2023cache}, denoted as BkPLLRead(LLC) and BkPLLWrite(LLC) for read and write, respectively, are particularly effective in influencing ATE. Specifically, the \textit{BkPLLRead(LLC)} attack increased the median ATE to over 1.7 (49X increase over solo), and the \textit{BkPLLWrite(LLC)} attack increased it to over 1.9 (55X increase). In simpler terms, these attacks caused OV$^2$SLAM's detected trajectory to deviate from the ground truth trajectory by a median value of two meters, and more than four meters in the worst case.
Both cache bank-aware DoS attacks are specially designed to generate many concurrent accesses to a specific LLC cache bank, causing contention on the bank. The shared L2 cache of Cortex-A57 consists of two tag banks, each composed of four data banks that can be accessed in parallel~\cite{bechtel2023cache}. By directing concurrent accesses to a single data bank, the SLAM task's access to the cache bank is delayed, subsequently delaying the execution of the SLAM task. 

\subsection{Impact of Co-scheduling HopeNet-Lite on Integrated GPU} \label{sec:eval-gpu}

In this experiment, we evaluate the impact of co-scheduling the HopeNet-Lite head pose estimator on the performance of the OV$^2$SLAM task. We compare the following configurations: \textit{Solo}, \textit{+DNN}, and \textit{+DNN\&DoS}. 
In \textit{Solo}, the OV$^2$SLAM runs alone; In \textit{+DNN}, the HopeNet-Lite DNN is co-scheduled with OV$^2$SLAM; In \textit{+DNN\&DoS}, both HopeNet-Lite and the DoS attack tasks are co-scheduled with OV$^2$SLAM. Note that the DoS attackers are best-effort tasks while both OV$^2$SLAM and HopeNet-Lite are real-time tasks. This implies that the DoS attackers cannot preempt OV$^2$SLAM or HopeNet-Lite.
% In particular, we add the DNN task~\cite{hopenetlite} to our tests. With the DNN task, we run two new test setups that both incorporate the GPU-based task: \textit{+DNN}, where only the OV$^2$SLAM and DNN tasks run, and \textit{+DNN\&DoS}, where OV$^2SLAM$ is run alongside both the DNN task, and four best-effort \textit{BkPLLWrite(LLC)} CPU tasks. 
% Based on the results from Figure~\ref{fig:dos-boxplot}, 
Based on the findings in Section~\ref{sec:eval-dos}, we opt for the \textit{BkPLLWrite(LLC)} attack as the aggressor workloads as it is the most effective at degrading the performance of the SLAM task. %Unlike the DoS attackers, though, we run the DNN model with a real-time priority of 1, as the head pose estimation task is listed as a high priority task in the target AR-HUD application~\cite{andreozzi2022industrial}. In addition, we pin the HopeNet-Lite task to Core 3 so that none of its required CPU operations (e.g. CUDA kernel launch, etc.) run on the same core as the SLAM task.

% \begin{figure*}[ht]
%   \centering
%   \begin{subfigure}{\textwidth}
%     \includegraphics[width=\textwidth]{figs-v1/DnnGpuDoSATE.pdf}
%     \caption{ ATE }
%     \label{fig:gpuresults-ate}
%   \end{subfigure}
%   \begin{subfigure}{\textwidth}
%     \includegraphics[width=\textwidth]{figs-v1/frames-dropped.pdf}
%     \caption{ \% of Frames Processed }
%     \label{fig:gpuresults-frames}
%   \end{subfigure}
%   \caption{ OV$^2$SLAM performance alongside a GPU-based HopeNet-Lite model and CPU-based \textit{BkPLLWrite(LLC)} attackers. }
%   \label{fig:gpuresults}
% \end{figure*}

\begin{figure*}[htp]
  \centering
  \begin{subfigure}{0.45\textwidth}
    \includegraphics[width=\textwidth]{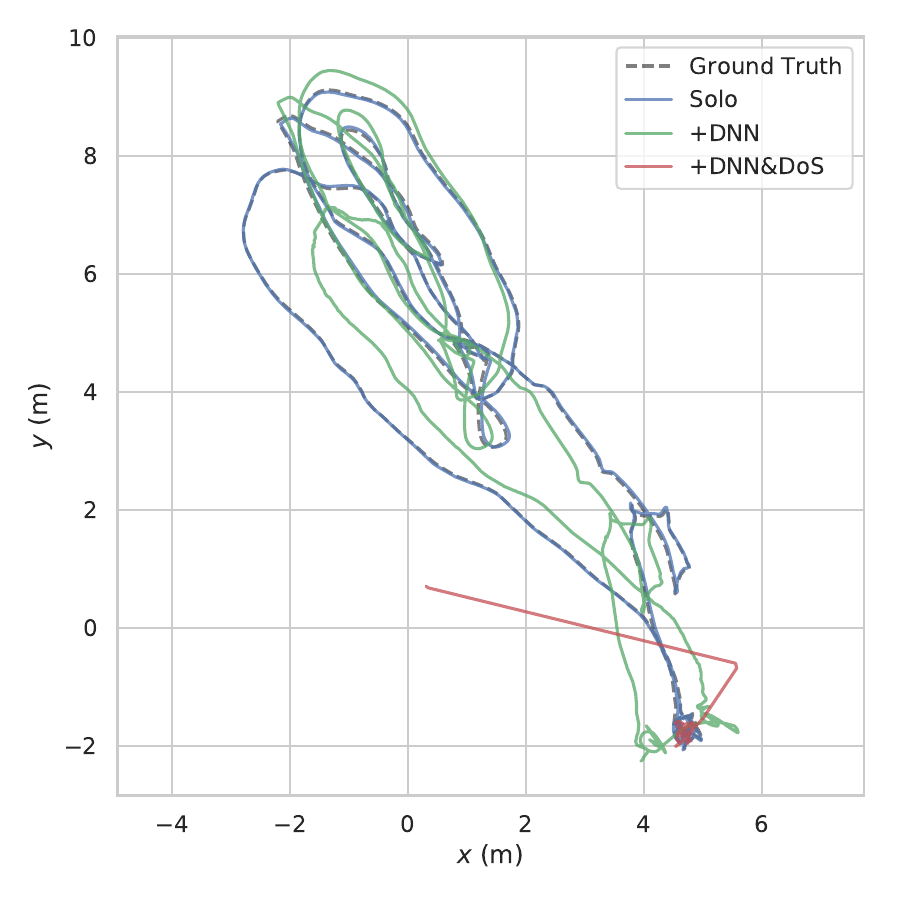}
    \caption{ Trajectory in XY plane }
    \label{fig:gpu-xy}
  \end{subfigure}
  \begin{subfigure}{0.45\textwidth}
    \includegraphics[width=\textwidth,page=2]{figs-v1/DnnTraj.pdf}
    \caption{ X, Y and Z positions over time }
    \label{fig:gpu-xyz}
  \end{subfigure}
  \caption{ OV$^2$SLAM trajectory when run alongside both \textit{BkPLLWrite(LLC)} DoS attackers and a GPU-based DNN application. }
  \label{fig:gpu-traj}
\end{figure*}

% \begin{figure}[htp]
%   \centering
%   \includegraphics[width=0.49\textwidth]{figs-v1/frames-dropped.pdf}
%   \caption{ Percentage of input frames processed by OV$^2$SLAM with resource intensive CPU and GPU co-runners. }
%   \label{fig:gpuresults-frames}
% \end{figure}

Figure~\ref{fig:gpu-traj} shows the ground-truth and the three generated trajectories. First, in \textit{Solo}, the generated trajectory almost completely overlaps with the ground-truth, indicating OV$^2$SLAM achieves good accuracy, with a median ATE of 0.03m, meaning the observed trajectory error is only about 3 centimeter.
In \textit{+DNN}, however, the addition of the HopeNet-Lite DNN task significantly impacts the accuracy of OV$^2$SLAM with the median ATE increasing to over 0.8m (approximately 27X increase over solo). 
This is because the OV$^2$SLAM and the HopeNet-Lite DNN task compete for shared hardware resources, particularly the shared DRAM, which is shared between the CPU (executing the SLAM) and the integrated GPU (executing the DNN).
Lastly, in \textit{+DNN\&DoS}, when HopeNet-Lite is combined with the DoS attacks, OV$^2$SLAM suffers a drastic performance degradation, leading to a complete failure in generating a full trajectory. This failure is due to  OV$^2$SLAM's inability to keep up with the input data, resulting in dropping a majority of the input camera frames. 
Our analysis shows that only about 26\% of the input image frames were processed in \textit{+DNN\&DoS}, compared to over 97\% in either the \textit{Solo} or \textit{+DNN} cases. 
Once again, contention on shared hardware resources among the co-scheduled tasks, particularly on the shared DRAM (caused by the DNN task) and the LLC bank (caused by the DoS attacks), contributes to this performance degradation.
%When the DoS attackers are added the median ATE does decrease, but the SLAM performance is instead degraded in an entirely different manner. 

In summary, co-scheduling HopeNet-Lite DNN and/or BkPLLWrite(LLC) DoS attacks significantly affects the accuracy of OV$^2$SLAM due to contention on shared hardware resources, despite the CPU and cache space being partitioned to protect the OV$^2$SLAM task.

\subsection{Runtime Analysis of OV\texorpdfstring{$^2$}SSLAM and HopeNet-Lite} %Not sure why, but two S are needed for one to appear in actual subsection header

To further investigate the impacts of shared resource contention on the AR-HUD application, we perform a detailed execution time analysis on the three CPU threads of the OV$^2$SLAM task---the \textit{Front End}, \textit{Mapping} and \textit{State Optimization} threads---and the DNN-based head pose estimator HopeNet-Lite. 

For OV$^2$SLAM (visual SLAM), we measure and record the execution times of each thread of the task when they execute their main computational loop (e.g. \textit{Front End} receives a new input frames, \textit{Mapping} receives a new keyframe, etc.). We re-run the experiments on the following three scenarios: \textit{Solo} when the SLAM task and its three threads run alone; \textit{+DoS} when the co-schedule the \textit{BkPLLWrite(LLC)} DoS attack tasks alongside with the SLAM; and \textit{+DNN} when we co-schedule the HopeNet-Lite task. 

For HopeNet-Lite (head pose estimation), we measure its inference times across 1000 input frames. We then compute the distribution of execution times for all tasks and threads to determine whether any of them experience execution delays due to co-runner interference. We re-run the experiments on the following three scenarios: \textit{Solo} when it runs alone, \textit{+SLAM} when run with the OV$^2$SLAM task, and \textit{+SLAM\&DoS} when run with both the OV$^2$SLAM task and \textit{BkPLLWrite(LLC)} DoS attack. 

\begin{figure*}[htp]
  \centering
  \begin{subfigure}{0.35\textwidth}
    \includegraphics[width=\textwidth]{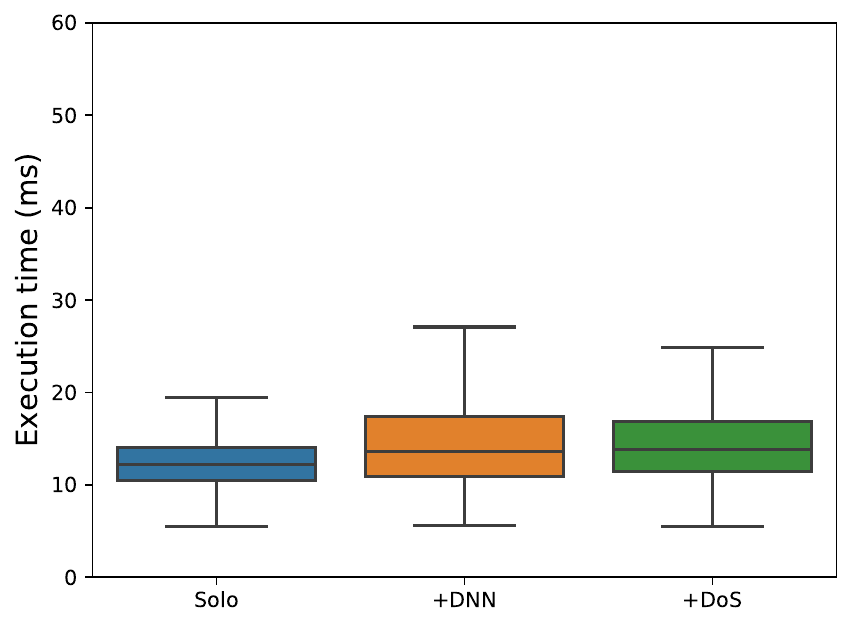}
    \caption{ Front End } % (3651 samples) 
    \label{fig:frontend-opt}
  \end{subfigure}
  \begin{subfigure}{0.35\textwidth}
    \includegraphics[width=\textwidth]{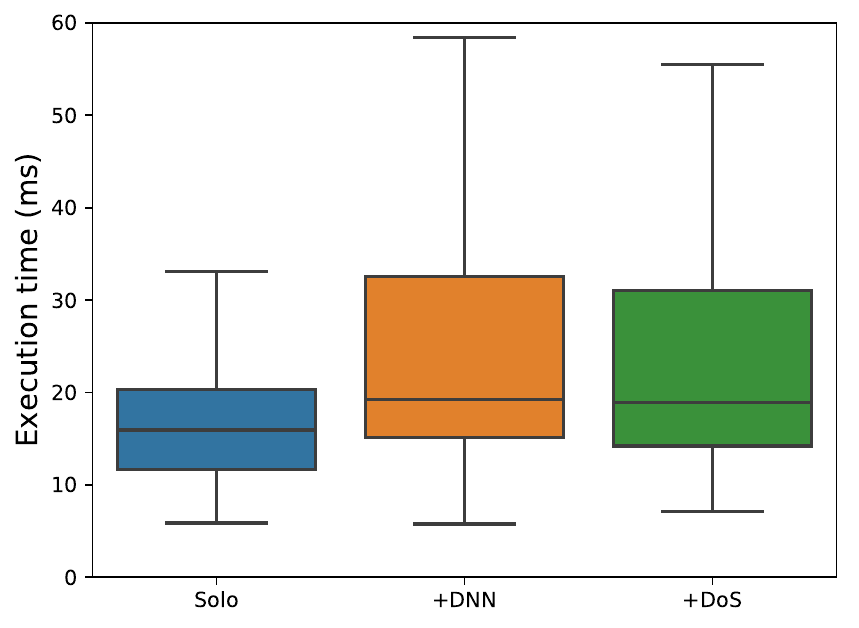}
    \caption{ Mapping  } % (581 samples)
    \label{fig:mapper-opt}
  \end{subfigure}
  \begin{subfigure}{0.35\textwidth}
    \includegraphics[width=\textwidth]{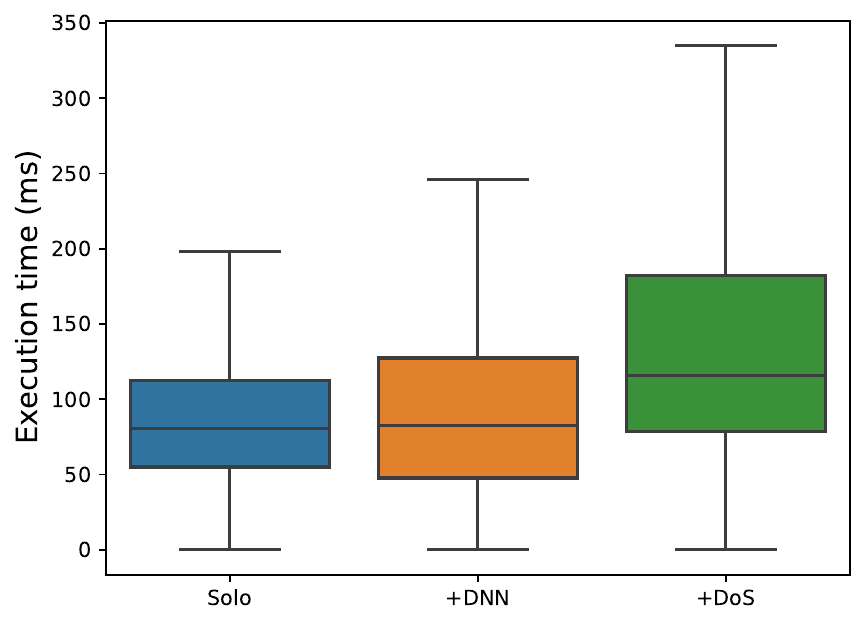}
    \caption{ State Optimization} %  (581 samples) 
    \label{fig:stateopt-cdf}
  \end{subfigure}
  \begin{subfigure}{0.35\textwidth}
    \includegraphics[width=\textwidth]{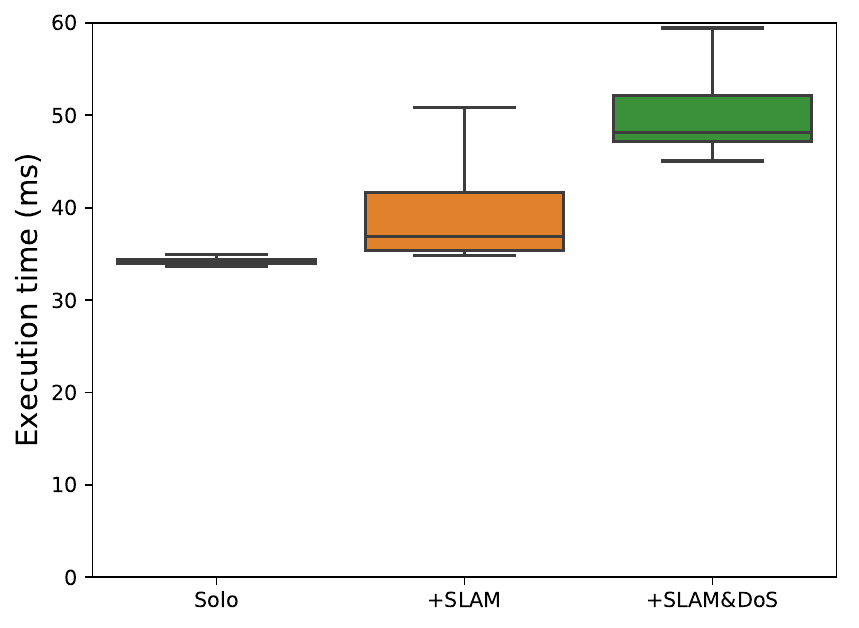}
    \caption{ Head Pose Estimation  } % (1000 samples)
    \label{fig:headpose-cdf}
  \end{subfigure}
  \caption{ Execution time distributions of all real-time tasks of the AR-HUD: (a)-(c) OV$^2$SLAM; (d) HopeNet-Lite}
  \label{fig:ov2slam-cdfs}
\end{figure*}

Figure~\ref{fig:ov2slam-cdfs} shows the execution time distributions for all real-time tasks (OV$^2$SLAM and HopeNet-Lite DNN) and their threads in each of the tested scenarios. 
First, for the OV$^2$SLAM task, we observe execution time increases in all three threads when co-runners (the HopeNet-Lite DNN or DoS attacks) are present. In case of the \textit{Front End} thread, inset (a), the execution time increases due to either type of the co-runners are relatively small. In case of the \textit{Mapping} thread, inset (b), both the DNN and the DoS attackers significantly increase the mapping thread's execution time. In case of the \textit{State Optimization} thread, inset (c), on the other hand, the DoS attackers are mo`re effective than the DNN co-runner in increasing its execution time. Recall that the DNN and the DoS tasks cause contention on different shared hardware resources, namely DRAM (bandwidth) and LLC (bank), respectively. As such, we can infer that (1) the Front End thread is not very sensitive to either of the shared resources, (2) the Mapping thread is sensitive to both DRAM and LLC, and (3) the State Optimization thread is more sensitive to the LLC. 
% Notably, though, we find that the \textit{Front End} and \textit{Mapping} threads see greater WCET slowdowns from the DNN co-runner than they do the DoS attacks. On the other hand, the \textit{State Optimization} thread is much more impacted by the DoS attacks in terms of the observed WCET. 
These execution time increases, especially in Mapping and State Optimization threads, due to contention result in the ATE increases we observe when the DNN or the DoS attack tasks are present. Note that the observed ATE loss is much higher when the DoS attackers are present, compared to when only the DNN co-runner is present (55X vs. 27X ATE increase over Solo), suggesting the importance of State Optimization  in the overall accuracy of the OV$^2$SLAM algorithm, 
% Given the disparity in SLAM ATE loss between the DoS attackers (55X) and the DNN co-runner (27X), we find that \textit{State Optimization} is of greater importance for OV$^2$SLAM in generating more accurate trajectories, 
which is consistent with prior findings~\cite{li2022timing}. 

Second, inset (d) shows the execution time distributions of the HopeNet-Lite DNN task. Note that when it runs together with the SLAM (on different cores/GPU), its average inference time increase from $\sim$34 ms (solo) to $\sim$37 ms (+SLAM). When the DoS attacks are then added (+SLAM\&DoS), however, the average inference time increases again to $\sim$48 ms. From these, we can infer that the DNN task itself is sensitive to both the SLAM task---especially with its Mapping thread, which may be contending DRAM bandwidth with the DNN---and the DoS attack task, which causes contention on a specific LLC bank. 
%From this, though, we find that the SLAM and DNNs task can run in real-time alongside each other when the DNN task is reduced to a control frequency of $\sim$20 Hz. As such, we use a control frequency of 20 Hz for the DNN task in all subsequent experiments. 
%we see larger amounts of delays in the \textit{Mapping} and \textit{State Optimization} threads, which is consistent with prior findings~\cite{li2022timing}. Even in the \textit{Front End} thread, we still see the WCET increase from $\sim$30 ms in the \textit{Solo} case to $>$100 ms in the \textit{DNN} case. From this, we do find execution time delays to likely be the cause for the performance loss observed in the OV$^2$SLAM task.

% Based on these results, we next explore a mitigation strategy for these adversary tasks.
% We next explore potential mitigation. 

\section{Mitigating Shared Resource Contention}~\label{sec:mitigation}

% \fixme{as described in the previous section, existing real-time scheduler in Linux alone fails to provide real-time performance guarantee due to shared resource contention. In this work, we use RT-Gang, which is a more advanced real-time scheduling framework, which enables bandwidth throttling of best-effort (non RT) tasks at the scheduler level. However, the baseline RT-Gang was originally designed for gang scheduling of a single parallel real-time task at a time where as our case-study applications need scheduling of two ...} \reply{I added this discussion below, starting in the second line.}

In this section, we present a mitigation solution to protect the real-time AR-HUD application in the presence of aggressor tasks (i.e., DoS attacks). 

\subsection{RT-Gang++}~\label{sec:rt-gang++}

In this work, we leverage the RT-Gang scheduling framework, which is a real-time gang scheduler, implemented as an extension to the \texttt{SCHED\_FIFO} real-time  scheduler in the Linux kernel~\cite{ali2019rt}.
RT-Gang supports a simple real-time gang scheduling policy, which allows only one parallel real-time gang at a time across all cores. 
Moreover, RT-Gang supports memory bandwidth throttling of best-effort tasks to protect any currently running real-time gang task. In other words, RT-Gang throttles any cores that execute best-effort tasks whenever a real-time gang task is running on any cores in the system. On the other hand, if no real-time gang is scheduled on the system, then the best-effort tasks have full access to the memory bandwidth. 

% challenges
When we tried to apply RT-Gang to mitigate the shared resource contention problem in the AR-HUD application of the ARM industrial challenge, we encountered the following challenges. 
First, the original RT-Gang supports only a single gang task at a time.
%While multiple tasks can be grouped together to form a ``virtual gang'' task~\cite{ali2021virtual}, it is only applicable when all tasks of the virtual gang have the same period and the same real-time priority. As described in Table~\ref{tbl:case-study-threads}, however, one of the three real-time tasks, the DNN task, in our case study has a lower real-time priority than the other two real-time tasks. 
When we schedule the real-time tasks as separate gang tasks in the original RT-Gang, they cannot meet the necessary real-time requirements because they are not fully parallelized, not taking advantage of all available cores and the GPU.  
Second, even if we group them together to form a ``virtual gang'' task~\cite{ali2021virtual} to improve resource utilization, RT-Gang does not offer any contention mitigation mechanisms between the real-time tasks within the virtual gang. This means that there is no way to minimize negative performance impact on a higher priority OV$^2$SLAM task due to co-scheduling the lower priority DNN real-time task, which runs on the iGPU. 
Lastly, when we deploy the cache bank-aware DoS attack~\cite{bechtel2023cache} as the best-effort aggressor tasks, RT-Gang's memory bandwidth throttling capability becomes ineffective in protecting the real-time tasks because the aggressors do not consume any memory bandwidth as they target LLC bank contention.

To address these challenges, we make three \textit{extensions} to the vanilla RT-Gang: 
(1) partitioned real-time gang scheduling capability;
(2) iGPU bandwidth throttling;
(3) LLC bandwidth throttling.
We call the resulting system \emph{RT-Gang++}. 
% utilizing the hardware-level GPU bandwidth throttling capability of Tegra X1 SoC's memory controller.

\subsection{Partitioned Gang Scheduling}

One major feature of the baseline RT-Gang is that it allows only one real-time gang task at a time across all the cores in a multicore CPU. While it does prevent shared resource contention between RT tasks by design, which is desirable for predictability, it is also a limitation in terms of scalability because not all tasks can benefit from a large number of cores and the number of cores in CPUs keeps increasing. While RT-Gang somewhat mitigates the problem by supporting so called ``virtual gangs'', which is a collection of multiple real-time tasks with the same period and the same priority that collectively acts like a single gang task, it cannot be used when either the priority or the period of any RT task differs from the rest. 
Moreover, many modern multicore CPUs are often composed of multiple clusters, each of which may have different sets of computing and memory resources that are not shared with the rest. 
For example, many ARM multicore CPUs incorporate the big.LITTLE architecture where one cluster is composed of powerful ``big" cores and the other cluster is composed of efficient ``little" cores. Not only do the clusters have different types of CPU cores, but they also often have cluster-private shared resources that are not shared across the clusters, which reduces the need to strictly adhere to the one-gang-task-at-a-time policy of RT-Gang. 

In RT-Gang++, we support multiple \emph{partitions} where each partition is composed of a statically determined set of cores. The one-gang-task-at-a-time is then applied to each partition rather then being applied globally. This means that multiple gang tasks, each with different priority and/or period, can run simultaneously as long as they are assigned to different partitions. For our AR-HUD case-study in Table~\ref{tbl:case-study-threads}, we assign OV$^2$SLAM and EuRoC playback tasks to form a virtual gang task and assign it into a gang partition, which is comprised of core 0, 1, and 2, while assigning the DNN task (head pose estimation) on another gang partition, which is composed of the core 3 and the iGPU. 
In this way, the system can run two active gang tasks with different priority and period simultaneously. 

\subsection{iGPU Memory Bandwidth Throttling} \label{sec:igputhrottle}

When multiple real-time tasks run together, however, they inevitably contend on the shared resources. As observed in Section~\ref{sec:eval-gpu}, co-scheduling the DNN task in particular has detrimental effect to the more critical OV$^2$SLAM task's accuracy. The baseline RT-Gang, unfortunately, does not provide any means to address the contention between the two co-scheduled RT tasks. In this work, we leverage hardware-level GPU bandwidth throttling capability of the platform we used for evaluation. Specifically, the Tegra X1 SoC of the Jetson Nano platform supports a number of QoS features at its memory controller, one of which is hardware-level throttling of subsets of hardware components that access the memory controller~\cite{nvidia2019}. The throttling feature of the memory controller provides 32 programmable throttling levels that can be applied to throttle the integrated GPU, which we used to throttle the DNN task whose memory access from the GPU impacts the performance of the OV$^2$SLAM task.

\begin{figure}[htp]
  \centering
  \includegraphics[width=0.45\textwidth]{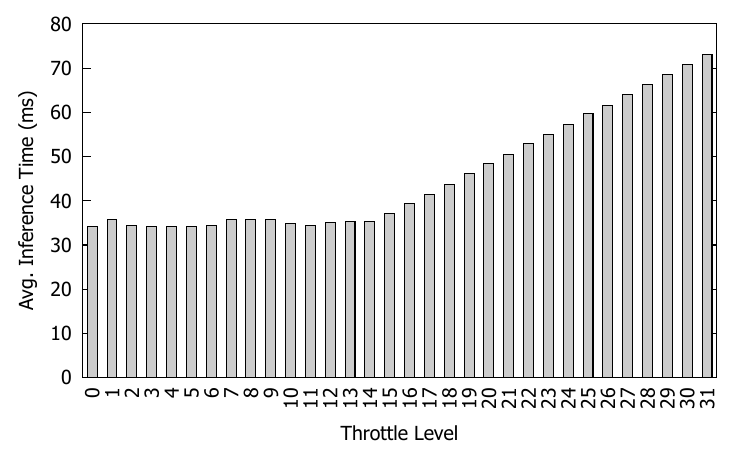}
  \caption{ Impact of GPU throttling on HopeNet-Lite DNN. }
  \label{fig:all-throttle}
\end{figure}

Figure~\ref{fig:all-throttle} shows the average HopeNet-Lite inference latencies under each of the 32 throttling levels when running alone in isolation. Note that GPU throttling does not impact the performance of the DNN task until the throttling level reaches around 15, indicating that sufficient bandwidth is provided until that point. 
% This indicates that the DNN model itself is also relatively insensitive to memory performance, as performance would otherwise be affected with GPU throttling enabled. 
After that, more aggressive throttling does impact the performance of the DNN. When the throttling level is 31, which is the maximum, the average inference latency is increased to $\sim$73 ms, which is about twice longer than without throttling. Note that there is a tradeoff between the accuracy of the SLAM task and the latency of the DNN task as more aggressive throttling of the GPU, which is used by the latter, will be helpful to achieve higher accuracy (lower ATE) for the SLAM task but it will increase the latency of the DNN task. As such, finding a ``sweet spot'' for the target application is necessary. For our study, we experimentally chose the level 20 as it was the maximum throttling level that still can provide 20Hz real-time performance for the DNN inference task. However, one can choose a more aggressive throttling level (e.g., the level 31) if achieving the highest accuracy of the SLAM task is more important than processing the DNN task at 20Hz. In our testing, using the GPU throttling level 31 allows the SLAM task to achieve near perfect isolation but at the cost of doubling the latency of the DNN task.

% , meaning that it can still run in real-time. Based on these findings, we keep the throttle level at 31 and reduce the periodic rate of the head-pose estimation task to 10 Hz. \fixme{this is not consistent with table II. is there a throttling level that can give good performance for both the ov2slam and the hopenet? for example, what happens when the gpu throttling level is 20?}

\subsection{LLC Bandwidth Throttling}\label{sec:llcthrottle}

For our case-study application, both the OV$^2$SLAM and DNN real-time tasks must be protected from the interference of co-scheduled aggressor tasks, which are scheduled in a best-effort manner (i.e., scheduled on any cores that do not execute RT tasks). 
As discussed in Section~\ref{sec:eval-dos}, when DoS attackers are used as the aggressor tasks, the performance of the OV$^2$SLAM algorithm is significantly reduced even though the DoS attackers cannot preempt the SLAM task due to shared resource contention. In particular, we find that cache bank-aware DoS attack~\cite{bechtel2023cache} is particularly effective in negatively impacting accuracy of the SLAM task. Unfortunately, however, the baseline RT-Gang's memory bandwidth throttling capability does not provide any protection against the cache-bank DoS attack, as it generates LLC cache hits and does not consume any memory bandwidth. 

In RT-Gang++, we add support for LLC bandwidth throttling capability by utilizing the L1-D cache miss performance counter (\texttt{L1D\_CACHE\_REFILL}) of the CPU cores to track and throttle LLC (L2) bandwidth used by the best-effort DoS attacker tasks. Note that the throttling implementation is based on MemGuard~\cite{yun2013rtas} and we use an experimentally determined LLC bandwidth threshold of 100 MB/s when LLC throttling is enabled and the regulation interval is 1ms.

\begin{figure*}[htp]
  \centering
  \begin{subfigure}{0.45\textwidth}
     \includegraphics[width=\textwidth]{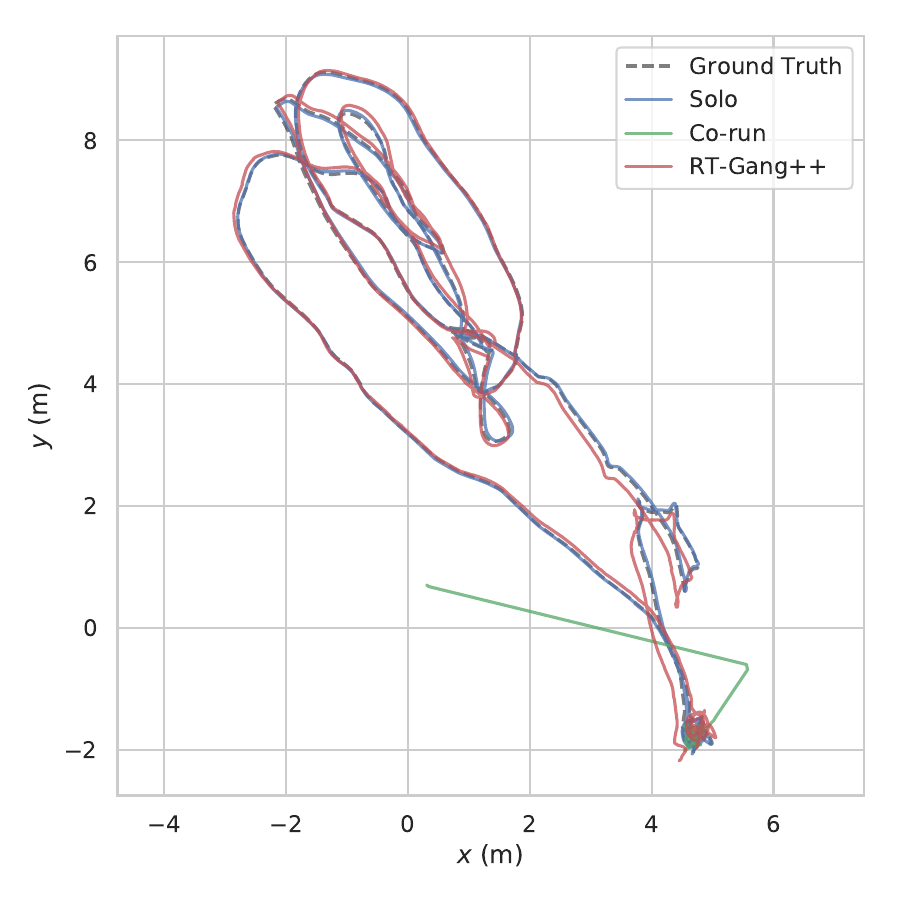}
     \caption{ Trajectory in X-Y plane }
     \label{fig:throttle-traj-plot}
  \end{subfigure}
  \begin{subfigure}{0.45\textwidth}
     \includegraphics[width=\textwidth,page=2]{figs-thesis/throttle-trajectory.pdf}
     \caption{ X, Y, and Z positions over time. } 
     \label{fig:throttle-traj-xyz}
  \end{subfigure}
  \caption{ Impact of RT-Gang++ on OV$^2$SLAM performance on the Jetson Nano. 
  % \fixme{can you change the color of Co-run to red (same as in Fig 3 '+DNN&DoS')
  %\fixme{legends: ground, solo, +dnn\&dos, +dnn\&dos (rt-gang++)} \reply{I updated the figure with new legends} 
  %\fixme{+DNN+Dos(RT-Gang++) --> RT-Gang++} \reply{Done}
  } % +dnn&dos (rt-gang++ w/o gpu throttling);
  \label{fig:throttle-traj}
\end{figure*}

\section{Evaluation Results}~\label{sec:results}

In this section, we evaluate the performance of RT-Gang++ on two popular embedded multicore platforms: Jetson Nano and Raspberry Pi 4.

\subsection{Jetson Nano}

The basic experiment setup is the same as described in Section~\ref{sec:case}: that is, we execute the SLAM task on Cores 0 and 1, the EuRoC dataset playback task on Core 2, and the DNN task on Core 3 and the iGPU. 

For evaluation, we repeat the experiment in Section~\ref{sec:eval-gpu} but with using RT-Gang++. Recall that in this experiment, we co-schedule BkPLLWrite(LLC) DoS attackers as aggressors on all CPU cores to incur contention on the shared hardware resources, specifically on a single LLC bank. As such, we want to know how well RT-Gang++ can protect the performance of the SLAM and the DNN task in the presence of the DoS attackers. 
\subsubsection{Results}

Figure~\ref{fig:throttle-traj} shows the trajectories generated by the OV$^2$SLAM task, and the X, Y and Z positions of the trajectory over time. 
Note that \textit{Co-run} denotes the baseline configuration without using RT-Gang++, while \textit{RT-Gang++} denotes the same configuration with RT-Gang++ enabled.
%\fixme{why 6(a) co-run is slightly different from 3(a)? is it due to run-to-run variation? if so, we need to use the exactly same data for both figures. this point is raised by the reviewer. So, we should carefully and clearly answer this. this is important.} \reply{I forgot one of the scaling options when plotting Figure 3 originally, so that caused it to be slightly different from Figure 6. I replotted and updated Figure 3 with the correct options so it should now match Figure 6.}
As observed earlier, in \textit{Co-run}, the OV$^2$SLAM fails to generate valid trajectory due to contention. In \textit{RT-Gang++}, however, the generated trajectory of the OV$^2$SLAM is valid and much closer to that of the \textit{Solo} and ground-truth, showing the effectiveness of the RT-Gang++ in protecting the performance of the SLAM task.

% Compared to \textit{Co-run}, where OV$^2$SLAM could only create a highly inaccurate partial trajectory, RT-Gang++ effectively protects against the \textit{BkPLLWrite(LLC)} attackers such that OV$^2$SLAM can produce a trajectory that is significantly closer to the ground truth. 
% Likewise, Figure~\ref{fig:rtgang-results-frames} shows the percentage of frames processed with our extended version of RT-Gang enabled. 
% Note that the OV$^2$SLAM is able to process $>$96\% of input frames with RT-Gang. 

% % However, we still find performance to be lacking with a median ATE $>$0.4, likely due to interference from the GPU. From this, we believe that more work can be done with regards to GPU-based mitigation strategies, which we leave to future work.

% \begin{figure}[htp]
%     \centering
%     \includegraphics[width=0.49\textwidth]{figs-v1/frames-dropped-rtgang.pdf}
%     \caption{ Percentage of input frames processed by OV$^2$SLAM with and without our extended version of RT-Gang enabled. }
%     \label{fig:rtgang-results-frames}
% \end{figure}

\begin{table}[htp]
  \centering
  % \begin{adjustbox}{width=0.49\textwidth}
  \begin{tabular}{|c|c|}
    \hline
    Configuration & Avg. inference time (ms) \\
    \hline
    Solo & 34.23 \\
    \hline
    Co-run  & 36.32 \\
    \hline
    RT-Gang++ & 48.69 \\ 
    \hline
  \end{tabular}
  % \end{adjustbox}
  \caption{Average inference latency of the DNN task }
  \label{tbl:dnn-inf}
\end{table}

% Likewise, we also measure RT-Gang++'s impact to the head pose estimation DNN task. 
Table~\ref{tbl:dnn-inf} shows the average inference times achieved by the HopteNet-Lite DNN task on three different test configurations. In \textit{Solo}, the DNN task runs alone in isolation and achieves an average inference latency of $\sim$34ms per image frame. In \textit{Co-run}, the SLAM and the DoS attackers are co-scheduled with the DNN, which results in a slight increase, going from $\sim$34 to $\sim$36 ms on average. In \textit{RT-Gang++}, on the other hand, the average latency of the DNN task is increased to $\sim$49 ms. This is because iGPU throttling limits the iGPU's DRAM bandwidth usage, which makes the DNN task runs slower, which in turn help protect the performance of the SLAM task, at the cost of the DNN. Nevertheless, it is important to note that the DNN task is still able to achieve the desired 20Hz rate (per Table~\ref{tbl:case-study-threads}).

% In the \textit{Co-run} case, we found that the DNN task's inference time was slightly impacted, going from $\sim$34 to $\sim$36 ms on average but that it could still meet the 50 ms deadline with ease. Similarly, with RT-Gang++ and iGPU bandwidth throttling enabled, we found that the DNN task's inference time was further increased to $\sim$49 ms on average, meaning that it could still achieve 20 Hz performance on average.

\begin{table}[htp]
  \centering
  % \begin{adjustbox}{width=0.49\textwidth}
  \begin{tabular}{|c|c|c|c|}
    \hline
    Core & Solo & Co-run & RT-Gang++ \\
    \hline
    0 & 8241 & 2413 & 956 \\
    \hline
    1 & 8470 & 1304 & 792 \\ 
    \hline
    2 & 7643 & 5628 & 1211 \\ 
    \hline
    3 & 7496 & 156 & 119 \\
    \hline
  \end{tabular}
  % \end{adjustbox}
  \caption{Average LLC bandwidth (in MB/s) consumed by all best-effort \textit{BkPLLWrite(LLC)} attackers. 
  % \fixme{how do you calculate this? total consumed divide by total time?} \reply{These are the bandwidth values reported by the BkPLLWrite attackers}
  }
  \label{tbl:dos-bw}
\end{table}

% To determine RT-Gang++'s impact to best-effort tasks, we also measure the performance of the \textit{BkPLLWrite(LLC)} attackers both with and without RT-Gang++ enabled. 
%For this, we again run the \textit{Co-run} and \textit{RT-Gang++} cases but measure the LLC bandwidth consumed by the DoS attacks in both scenarios. 

Table~\ref{tbl:dos-bw} shows the average LLC bandwidth consumed by the best-effort DoS attackers on each core first alone in isolation, in \textit{Solo}, together with both SLAM and DNN real-time tasks without and with RT-Gang++, in \textit{Co-run} and \textit{RT-Gang++}, respectively. 
Recall that RT-Gang++ throttles LLC bandwidth of best-effort tasks to limit the LLC bank contention.
As a result, the average LLC bandwidth numbers of \textit{RT-Gang++} is much lower than that of \textit{Co-run} or \textit{Solo}, which is expected.
Nevertheless, it is important to note that these best-effort DoS attackers are still able to use a significantly higher LLC bandwidth that the set threshold of 100MB/s (which was chosen experimentally per Section~\ref{sec:llcthrottle}).
This is because, in RT-Gang++, the LLC bandwidth throttling is dynamically enabled only when there are currently scheduled real-time tasks in any of the CPU cores in the platform. When there are no active real-time tasks, which happen often as they can finish earlier than the deadlines, LLC bandwidth throttling is automatically disabled, which allows the best-effort DoS attackers to fully utilize the full LLC bandwidth without throttling, until any of the real-time tasks become active again. This is

\begin{table}[htp]
  \centering
  % \begin{adjustbox}{width=.49\textwidth}
  \begin{tabular}{|c|c|c|c|}
    \hline
    \multirow{2}{*}{Dataset} & \multirow{2}{*}{Solo} & RT-Gang++ & \multirow{2}{*}{RT-Gang++} \\
    & & (No iGPU throttling) & \\
    \hline
    MH01    & 0.03 & 0.36 & 0.11 \\ % The 25Hz case had a median ATE of 0.42
    \hline
    MH02    & 0.04 & 0.47 & 0.07 \\
    \hline
    MH03    & 0.08 & 0.24 & 0.20 \\
    \hline
    MH04    & 0.15 & 2.86 & 0.29 \\
    \hline
    MH05    & 0.12 & 0.80 & 0.23 \\
    \hline
  \end{tabular}
  % \end{adjustbox}
  \caption{ OV$^2$SLAM median ATE (in meters) on the Machine Hall scenarios from the EuRoC dataset. 
  % Note that we do not include the ATE for the \textit{+DNN\&DoS} test case as OV$^2$SLAM could not process the datasets in real-time, and only produced partial trajectories. 
 % \fixme{table 4 and 5 should use the same columns. just focus on median ATE so that we can compare w/ and w/o rt-gang++} \reply{The columns should now match the other tables}
  } 
  \label{tbl:mh-ate-results}
\end{table}

% Table~\ref{tbl:mh-frame-results} shows the percentage of frames processed by OV$^2$SLAM in each test case. Much like the MH01 dataset, we find that OV$^2$SLAM is unable to process the majority of input frames in the \textit{Co-run} case. When RT-Gang++ is enabled, though, we are again able to achieve real-time performance as OV$^2$SLAM processes $>$97\% of input frames across all datasets. 

% Finally, to further test RT-Gang++, we run the OV$^2$SLAM task on the other four machine hall EuRoC scenarios, MH02-MH05. 

Table~\ref{tbl:mh-ate-results} then shows the median ATE values of the OV$^2$SLAM generated trajectories on all five machine hall scenarios of the EuRoC dataset. 
Note that \emph{Co-run} is not included as OV$^2$SLAM fails to generate full trajectories due to contention.
Even with RT-Gang++, we still see notable increases in reported ATE values when the iGPU bandwidth throttling is not enabled. On the other hand, when both LLC and iGPU bandwidth throttling are enabled in \textit{RT-Gang++}, we observe significant improvements in ATE. The degree of improvements differs depending on the scenarios. For example, the MH04 scenario, which is among the hardest ones, is particularly sensitive to the iGPU throttling as the SLAM's ATE is whopping 2.86 meters without iGPU throttling, while it is only 0.29 meters with iGPU throttling. 
% In general, harder scenarios such as MH04 and MH05 (recall that the MH scenarios are sorted according to the difficulty; see Section~\ref{sec:case}) are more affected by the lack of iGPU throttling, suggesting that they are 
The results show that both iGPU throttling and LLC bandwidth throttling are important to protect the performance of the SLAM task. 
Note that an ATE value less than 0.3 meters is required for these scenarios~\cite{li2022timing}, which we are able to satisfy in all cases with RT-Gang++ fully enabled. 

\subsection{Raspberry Pi 4}

To demonstrate the generality of RT-Gang++, we additionally evaluate it on a Raspberry Pi 4 platform. 

\subsubsection{Hardware and Software Setup}

%a{\bf Hardware Platform:} 
The Raspberry Pi 4 features a Broadcom BCM2711 SoC, which includes a quad-core ARM Cortex-A72 CPU with a 1MB shared L2 cache, a VideoCore 6 GPU, and 4GB LPDDR4 SDRAM. Note that the Raspberry Pi 4's Cortex-A72 CPU cores are more advanced and powerful than the Jetson Nano's Cortex-A57 cores. The Pi 4's VideoCore 6 GPU, though, is weaker than the Nano's 128-core Maxwell GPU and lacks software framework support for GPU offloading. Table~\ref{tbl:pi4-platform} shows the hardware specification for the Raspberry Pi 4.

\begin{table}[htp]
  \centering
  % \begin{adjustbox}{width=.49\textwidth}
  \begin{tabular}{|c|c|c|}
    \hline
    Platform                & Raspberry Pi 4 \\ 
    \hline
    SoC                     & BCM2711  \\ 
    \hline
    \multirow{1}{*}{CPU}    & 4x Cortex-A72 @ 1.5GHz   \\      
    \hline
    GPU                     & VideoCore 6 (not used)\\
    \hline
    Shared LLC (L2)         & 1MB (16-way)\\ 
    \hline
    Memory (Peak B/W)       & 4GB LPDDR4 (25.6 GB/s) \\
    \hline
  \end{tabular}
  % \end{adjustbox}
  \caption{Raspberry Pi 4 hardware specifications.}
  \label{tbl:pi4-platform}
\end{table}

% \fixme{Add Raspberry Pi 4 hardware spec. Table 1 may include only the Jetson Nano as in the original submission} \reply{Done, I also removed the Pi 4 specs from the Nano table}

%{\bf Application Setup:} 
The basic application setup is the same as that of the Jetson Nano in Table~\ref{tbl:case-study-threads}, except that HopeNet-Lite DNN task runs entirely on the CPU due to the lack of DNN software support for the Pi 4's GPU.

% {\bf Operating System Setup:} 
For the software, the Pi 4 runs Raspberry Pi OS with Linux kernel 5.15 and is also patched with PALLOC~\cite{yun2014palloc} to support LLC space partitioning, as was the case for the Jetson Nano platform. We use the same 2/2 split LLC partitioning setup as we did on the Jetson Nano, although the size of each cache partition is halved as the Pi 4's L2 cache size is a half of Nano's. 

% {\bf RT-Gang++ on Pi4:}
Lastly, we ported RT-Gang++ on the Pi 4's Linux kernel, including partitioned gang scheduling and LLC bandwidth throttling capabilities. However, the iGPU throttling is not implemented because the iGPU is not used as noted earlier.

\subsubsection{Results}

% \fixme{Add figure 2 equivalent for Pi 4} \reply{Done, Fig 7 shows the pi 4 results}
First, we repeat the experiment in Section~\ref{sec:eval-dos} to understand the effect of various DoS attacks on the Pi 4 platform. 

\begin{figure*}[htp]
  \centering
  \includegraphics[width=0.73\textwidth]{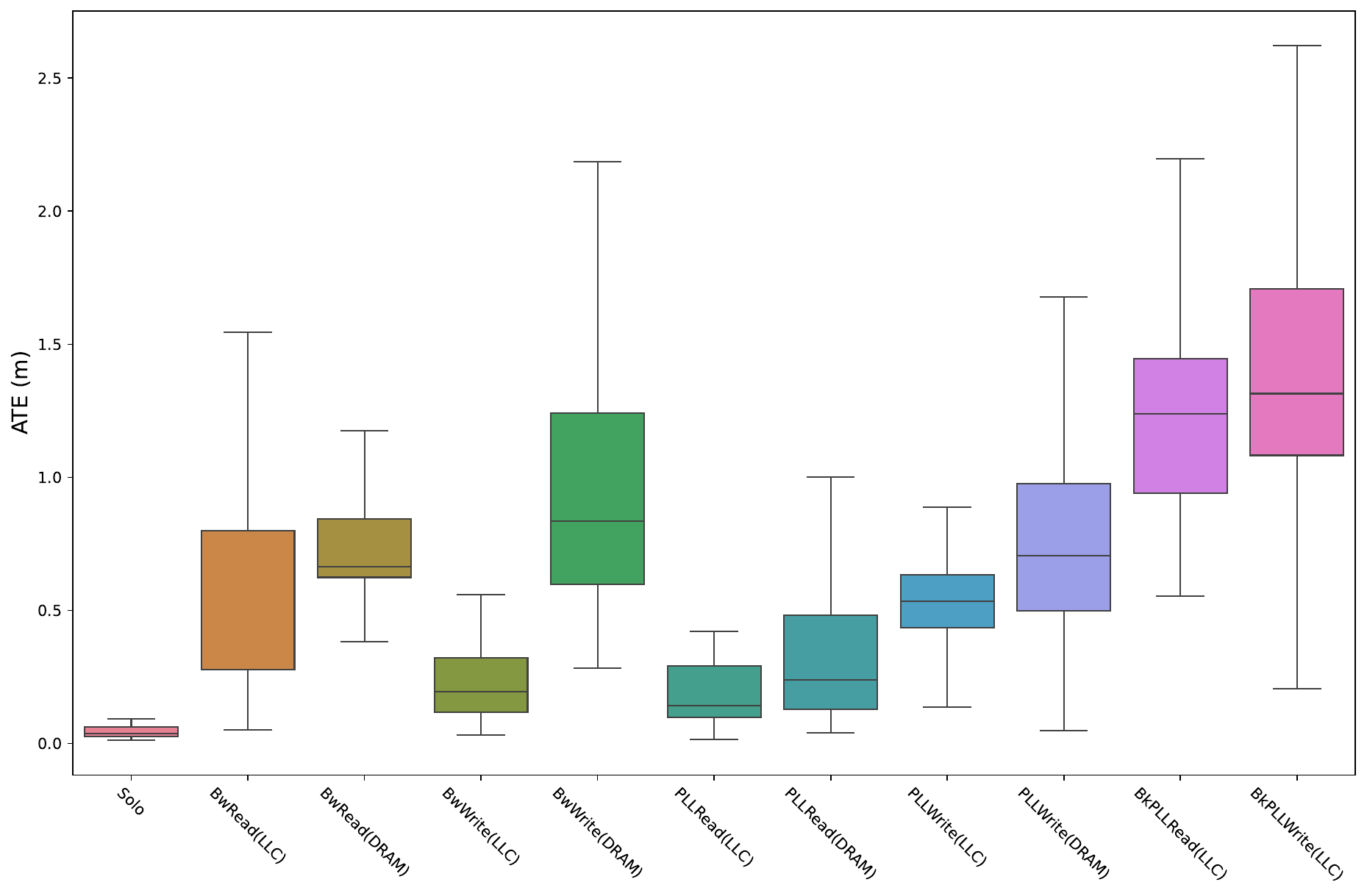}
  \caption{ Impact of DoS attacks on the Absolute Trajectory Error (ATE) of the OV$^2$SLAM generated trajectory on the Pi 4.} %  \fixme{font is too small}
  \label{fig:dos-boxplot-pi4}
\end{figure*}

Figure~\ref{fig:dos-boxplot-pi4} shows the results. Similar to the results on Jetson Nano in Figure~\ref{fig:dos-boxplot}, all DoS attacks increase ATE scores in the SLAM generated trajectories due to increased contention. Note also that BkPLLWrite(LLC) is again the most effective DoS attack, same as the Jetson Nano platform, although its median ATE increase of 1.3 is somewhat smaller than the median ATE of 1.9 we observe on the Jetson Nano platform. This suggests that Pi 4's Cortex-A72 CPU cores are also highly susceptible to LLC bank contention as in Nano's Cortex-A57. 
As such, we again use the BkPLLWrite(LLC) as the DoS attacker task in the subsequent experiments. 

% Note that we again use \textit{BkPLLWrite(LLC)} DoS co-runners in these experiments as we found them to  be the most effective on the Pi 4 as well. They caused the SLAM task's median ATE to increase to $\sim$1.3. 

Next, we repeat the experiment in Section~\ref{sec:eval-gpu} to evaluate the performance of RT-Gang++ in protecting the performance of the OV$^2$SLAM in the presence of the DoS attacks and the HopeNet-Lite DNN task. 

% We run the same test scenario as in Figure~\ref{fig:throttle-traj}. That is, we rerun the SLAM task alongside both the DNN and worst-case \textit{BkPLLWrite(LLC)} DoS attacks, but with RT-Gang++ enabled.

\begin{figure*}[htp]
  \centering
  \begin{subfigure}{0.45\textwidth}
     \includegraphics[width=\textwidth]{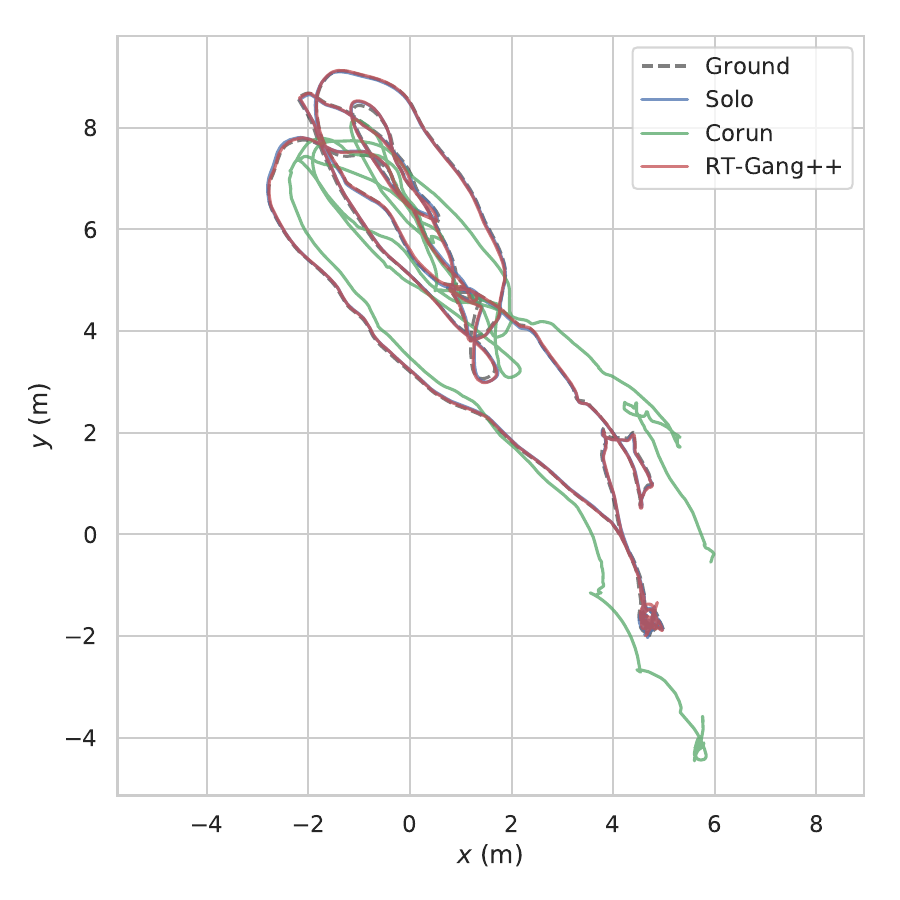}
     \caption{ Trajectory in X-Y plane }
     \label{fig:throttle-traj-plot-pi4}
  \end{subfigure}
  \begin{subfigure}{0.45\textwidth}
     \includegraphics[width=\textwidth,page=2]{figs-toc/Pi4-Fig6.pdf}
     \caption{ X, Y, and Z positions over time. } 
     \label{fig:throttle-traj-xyz-pi4}
  \end{subfigure}
  \caption{ Impact of RT-Gang++ on OV$^2$SLAM performance on the Pi 4. 
  % \fixme{can you change the color of Co-run to red (same as in Fig 3 '+DNN&DoS')
  %\fixme{legends: ground, solo, +dnn\&dos, +dnn\&dos (rt-gang++)} \reply{I updated the figure with new legends} 
  %\fixme{+DNN+Dos(RT-Gang++) --> RT-Gang++} \reply{Done}
  } % +dnn&dos (rt-gang++ w/o gpu throttling);
  \label{fig:throttle-traj-pi4}
\end{figure*}

Figure~\ref{fig:throttle-traj-pi4} shows the OV$^2$SLAM generated trajectories and XYZ positions over time with and without RT-Gang++ on the Pi 4 platform.
% Note that the trajectory labels indicate the same test scenarios as in Figure~\ref{fig:throttle-traj}. 
Note first that, like the results on the Jetson Nano platform (Figure~\ref{fig:throttle-traj}), 
we find that contention from the co-scheduled DoS attacks and the DNN task do impact performance of the SLAM. 
Unlike on Jetson Nano, however, the SLAM task is able to generate a full trajectory although the ATE is increased significantly and the generated trajectory in the \textit{Co-run} case is far from the ground-truth or that of the \textit{Solo} case. 
This is mainly because the DNN task is running on a CPU core on the Pi 4 instead of running on the GPU. On the Pi 4, the DNN task running on a CPU core does not generate as much DRAM bandwidth contention compared to if it was run on a GPU. 
% Note that the DNN task's average inference latency, when running alone in isolation, is around 300ms per frame, which is far slower than the required 50ms latency for 20Hz real-time operations. 
%To address this contention, we also employ RT-Gang++ on the Pi 4. 
Second, RT-Gang++ is able to effectively mitigate the contention generated by the DNN task and DoS attacks and protect the performance of the SLAM task. Concretely, RT-Gang++ reduces the median ATE down to only $\sim$0.05 meter, which is very close to the $\sim$0.04 median ATE seen in the \textit{Solo} case.
%\fixme{fix the number} \reply{Done}in the \textit{Solo}
%\fixme{Need to elaborate more, especially the impact of not using the gpu} \reply{I added more discussion of the DNN on CPU to the section above.}

% More importantly, we find that the DNN task can actually be beneficial for the SLAM task when DoS attacks are also present (without the DNN, the \textit{BkPLLWrite(LLC)} attackers increased the median ATE to $\sim$1.3). 
% This is because the DNN's inference time is much longer on the Pi 4 ($\sim$300 ms per frame), so the DNN model effectively utilizes 100\% of its assigned CPU core. 
% Consequently, this prevents the DoS attacker task on that core from ever running, meaning that less contention is generated as a result. 
% Even then, the SLAM task still sees a median ATE increase of $\sim$23X (0.92 meter).

\section{Related Work} \label{sec:related}

Simultaneous Localization and Mapping (SLAM) algorithms are at the heart of many robotics applications, including ADAS and self-driving systems, as they are used to localize the position and pose of the ego-vehicle in connection with the surrounding environment. 
% Due to the relatively cheaper cost of camera sensors, many efforts have been made towards solving the challenge of Visual SLAM. 
% The PTAM algorithm was the first to introduce a multi-threaded Visual SLAM implementation~\cite{klein2007parallel} for improved real-time performance. It separated tracking and mapping into separate parallel threads on a dual-core processor, which allowed them to create more detailed 3D maps and achieve state-of-the-art accuracy. This type of multi-threaded approach has since been widely adopted in many Visual SLAM implementations. 
Notable SLAM algorithms include the LSD-SLAM algorithm~\cite{engel2014lsd}, the many iterations of the ORB-SLAM algorithm~\cite{mur2015orb,mur2017orb,campos2021orb}, and the OV$^2$SLAM algorithm~\cite{ferrera2021ov}, which was used in this paper. These algorithms detect features of the camera images and track the features to locate their positions in the world. 
Recently, Li et. al, observed that performance of SLAM algorithms can be sensitive to execution timing delays and proposed an adaptive strategy within the SLAM to minimize performance degradation~\cite{li2022timing}. In contrast, our work proposes a system-level solution that does not require changes in the SLAM algorithm and provides in-depth microarchitectural analysis on resource contention.

% Other SLAM algorithms instead combine both camera and inertial sensors (e.g. IMU) to better account for visual artifacts such as low textures and motion blur. Better known as Visual-Inertial SLAM, or VI-SLAM for short, these algorithms allow for better robustness in situations where standard Visual SLAM algorithms may otherwise struggle. VI-SLAM approaches, though, tend to suffer from mid and long-term data association of generated 3D maps, meaning that the maps may not accurately reflect the surrounding environment if they are not updated frequently enough. To address this, the ORB-SLAM-VI algorithm used IMU preintegration to better allow for variable length data association~\cite{mur2017visual}. The ORB-SLAM3 algorithm then extended upon this by using MAP estimation to better account for sensor uncertainties~\cite{campos2021orb}. 
% In this work, however, we focus on a Visual SLAM algorithm, OV$^2$SLAM, in our case study as it is recommended for the target AR-HUD application.

Microarchitectural DoS attacks are software attacks specifically designed to induce a high-degree of resource contention. DoS attacks on several shared resources have been studied and evaluated. Moscibroda et al. proposed a DoS attack that targets a FR-FCFS scheduling algorithm~\cite{rixner2000memory} in shared DRAM controllers~\cite{moscibroda2007memory}. 
%As a result, many fair scheduling algorithms were introduced~\cite{mutlu2007stall,kim2010thread}. 
Keramidas et al. demonstrated DoS attacks targeting shared LLC space and, to address them, proposed a cache replacement policy that gave the attackers access to less of the LLC space~\cite{keramidas2006preventing}. Woo et al. investigated DoS attacks on bus bandwidth and shared cache space in a simulated environment~\cite{woo2007analyzing}. Valsan et al. and Bechtel et al. showed that DoS attacks could target internal LLC hardware structures~\cite{bechtel2019dos,bechtel2018picar,valsan2016taming}. Based on~\cite{bechtel2019dos}, Iorga et al. presented a statistical approach for testing DoS attacks~\cite{iorga2020slow}. Li et al. applied machine learning to better optimize DoS attackers, resulting in WCET slowdowns $>$400X~\cite{li2022polyrhythm}. GPU-based DoS attacks have also been studied by researchers. Yandrofski et al, systematically studied shared resource contention on discrete Nvidia GPUs by generating various adversarial programs~\cite{yandrofski2022making}. Bechtel et al., implemented DoS attacks targeted towards Intel iGPUs, as they also access the LLC. %studied potential DoS attacks on Intel iGPUs~\cite{bechtel2022denial}, which share a LLC with the CPU.

Much effort has been devoted to address the problem of shared resource contention in multicore in the real-time systems research community. 
%% suggested storyline
% partitioning -> cache, dram bank.
% throttling --> dram bandwidth, cache bandwidth
Partitioning of shared resources, especially shared cache~\cite{mancuso2013rtas,kim2013coordinated,kim2017attacking,farshchi2018deterministic,xu2019holistic,roozkhosh2020potential} and DRAM banks~\cite{kim2017attacking,farshchi2018deterministic}, has been extensively studied. 
Bandwidth throttling~\cite{yun2013rtas,xu2019holistic,ewarp20,saeed2022memory,saeed2023memory,MemPol23,eric2023bandwatch} has been another popular approach.
MemGuard~\cite{yun2013rtas} uses per-core hardware performance counters to throttle each core's bandwidth usage, which has been a standard throttling technique in many subsequent studies. 
These mechanisms are used to enable tighter worst-case timing analysis on multicore. An exhaustive review on multicore timing analysis can be found in~\cite{maiza2019survey}.
% MemPol proposed to use a dedicated core for more fine-grained and lower overhead throttling in some capable ARM platforms~\cite{MemPol23}.
Recently, both Intel and ARM also introduced hardware support for shared resource partitioning and throttling~\cite{intel-rdt,arm-mpam-supp}, though their effectiveness in providing isolation for real-time systems is still insufficient~\cite{zini2022analyzing,sohal2022closer,bechtel2023cache}. For example, it was reported that even at the maximum throttling level, Intel RDT was not able to protect critical real-time task because the throttled cores were still able to generate significant traffic, enough to cause a 80\% performance degradation of the real-time task in the worst-case~\cite{sohal2022closer}. Nevertheless, these hardware capabilities orthogonal and can easily be integrated into our framework.

In most of these works, cores are partitioned between RT and best-effort cores. 
For example, Saeed et al. proposed a memory utilization based dynamic throttling system that protects a single RT core by throttling memory bandwidth usage of the other best-effort cores~\cite{saeed2022memory,saeed2023memory}. Seals et al., also proposed a dynamic throttling system, which also can throttle iGPU's memory bandwidth, to protect a single RT core~\cite{eric2023bandwatch}.
However, such an approach can significantly under-utilize computing resources. For example, in the AR-HUD industrial challenge problem, one RT core is not sufficient as the application consists of a multi-threaded CPU task and a GPU task, both of which need real-time guarantees. 
RT-Gang~\cite{ali2019rt,ali2021virtual} offers more flexible scheduling potential as all cores can be utilized for both real-time and best-effort tasks. This is because the OS automatically throttles the best-effort tasks only when a RT task is running on any core(s) in the multicore system. However, it does not offer any protection between the real-time tasks on different cores and the iGPU. % does not provide throttling capability in novel cache bank attack, 
In this work, we leverage RT-Gang but address its limitations by adding iGPU throttling, CPU cache bandwidth throttling, and partitioned gang scheduling, which allowed us to successfully consolidate the ARM industrial challenge application while ensuring its performance in the presence of fully loaded  aggressors on a real heterogeneous multicore platform. 

\section{Conclusion} \label{sec:conclusion}

% In this paper, we perform a systematic case study on a real-world representative AR-HUD application. In the process, we find that shared resource contention is a serious concern with regards to SLAM algorithm performance. Furthermore, we show that the presence of resource intensive CPU and GPU-based co-runners can ultimately slowdown a SLAM task to the point where it can not process input sensor data in real-time. To address this, we employ an extended version of the RT-Gang scheduling framework that supports %partitioned gang scheduling and 
% LLC bandwidth throttling. In doing so, we are able to protect the SLAM task such that it can run in real-time again. However, we still see impacts to performance from GPU-based co-runners, so additional efforts will likely need to be made to guarantee isolation from the GPU.

In this paper, we presented a solution to the Industrial Challenge problem put forth by ARM in 2022~\cite{andreozzi2022industrial}. We systematically analyzed the effect of shared resource contention to an augmented reality head-up display (AR-HUD) case-study application of the industrial challenge on a heterogeneous multicore platform.
Using micro-architectural denial-of-service (DoS) attacks as aggressor tasks of the challenge, we showed that such aggressors can dramatically impact the latency and accuracy of the AR-HUD application, which could result in significant deviations of the estimated trajectories from the ground truth, despite the best effort to mitigate their influence by using cache partitioning and real-time scheduling of the AR-HUD application. To address this we propose RT-Gang++, which combines LLC and iGPU bandwidth throttling to mitigate shared resource contention from their respective resources. By deploying RT-Gang++, we were able to effectively protect the performance of the critical SLAM task, such that it could achieve near solo case performance, without having to over-provision the system.

% We showed that both selective LLC bandwidth throttling of the aggressor tasks and iGPU bandwidth throttling of the DNN task, combined, can be effective means to ensure real-time performance of the AR-HUD application without resorting to over-provisioning the system. 

For future work, we plan to perform similar case studies on more capable platforms than the Jetson Nano, such as the Jetson Xavier or Jetson Orin lines of embedded platforms, to evaluate their susceptibility to shared resource contention and the generality of the proposed RT-Gang++ framework.

% and better evaluate our RT-Gang++ framework on more complex platforms.
%-------------------------------------------------------------------------
\section*{Acknowledgements} \label{sec:acknowledge}

This research is supported in part by NSF  CNS-1815959, CPS-2038923 and NSA Science of Security initiative contract no. H98230-18-D-0009.

\bibliographystyle{abbrv}
\bibliography{reference}

\begin{IEEEbiography}
[{\includegraphics[width=1in,height=1in,clip]{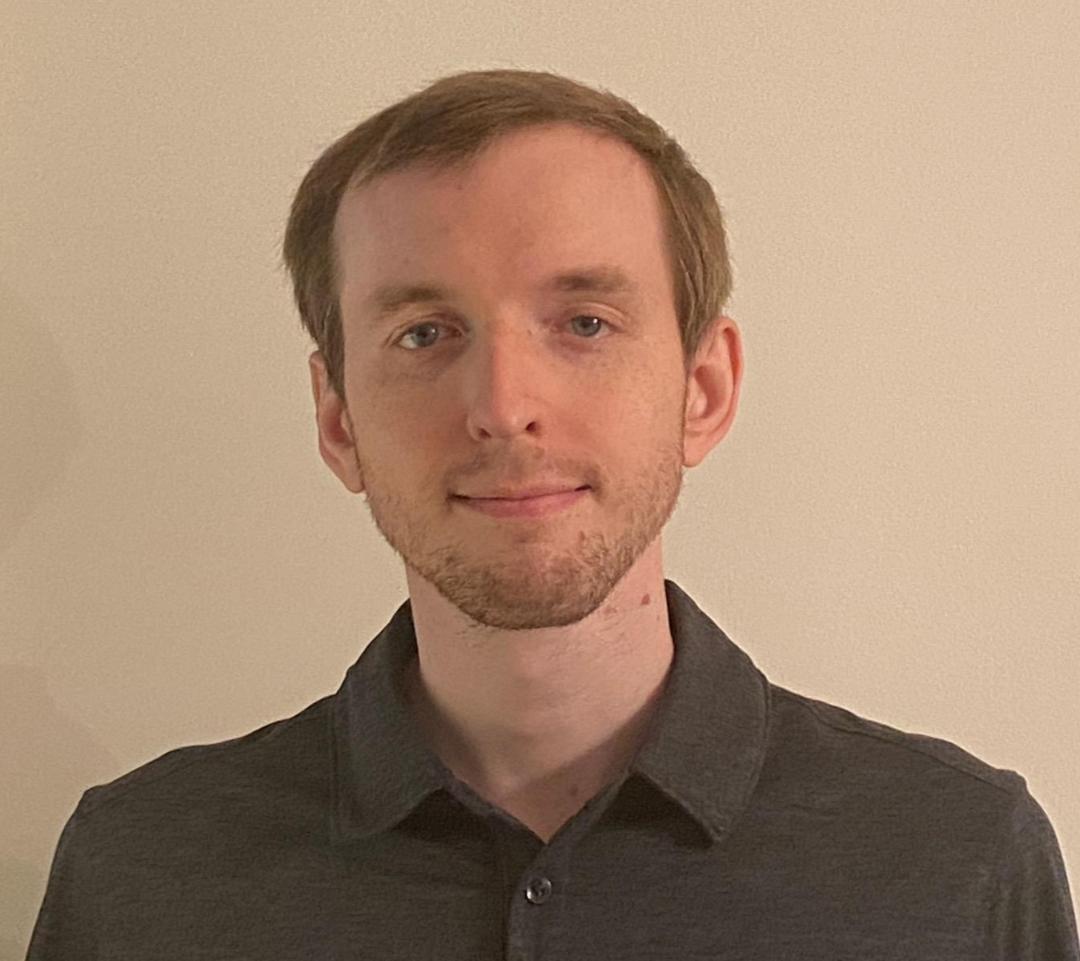}}]
{Michael Bechtel}

received a B.S. degree in Computer Science from the University of Kansas in 2017 and a Ph.D. degree in Computer Science from the University of Kansas in 2023. His research interests include real-time embedded systems, computer architecture, and cyber-physical systems. His work has appeared in top embedded real-time systems venues such as RTAS, and he received an Outstanding Paper Award from RTAS'19. He is currently a software engineer at Garmin.

\end{IEEEbiography}

\begin{IEEEbiography}
[{\includegraphics[width=1in,height=1.25in,clip,keepaspectratio]{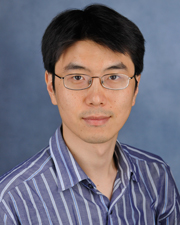}}]
{Heechul Yun}
is an associate professor in the department of Electrical Engineering and Computer Science at the University of Kansas. His research interests include OS, computer architecture, and real-time embedded systems with special emphasis on addressing real-time, security, and safety related issues on safety-critical cyber-physical systems (e.g., autonomous cars and UAVs). His work has appeared in top embedded real-time systems venues such as RTAS, ECRTS and Transactions on Computers; and received multiple prestigious paper awards (Best Paper Award from RTSS'20, Outstanding Paper Award from RTAS'19, Best Paper Award from RTAS'16, Editor's Pick of the Year Award from IEEE Transactions on Computers in 2016). He received a Ph.D. degree in Computer Science from the University of Illinois at Urbana-Champaign in 2013. Prior to his Ph.D., he worked at Samsung Electronics as a senior software engineer.

\end{IEEEbiography}

\end{document}